\documentclass[11pt]{article}
 \pdfoutput=1 
\usepackage[left=0.8in,right=0.8in,a4paper]{geometry}
\usepackage{empheq}
\usepackage{amssymb}
\usepackage[pdfstartview=FitH,pdfpagemode=None]{hyperref}
\usepackage{color}
\usepackage{slashed}
\usepackage{amsmath}
\usepackage{amsfonts}
\usepackage{amssymb}
\usepackage{authblk}
\usepackage{graphicx}
\usepackage{caption}
\usepackage{subcaption}
\usepackage{float}
\usepackage{tikz}
\usepackage{amsmath}
\usepackage{appendix}
\usepackage{listings}

\newcommand{\be}{\begin{equation}}
\newcommand{\ee}{\end{equation}}

\newcommand{\tr}{\text{tr}}
\def\bea{\begin{eqnarray}}
\def\eea{\end{eqnarray}}

\def\nn{\nonumber}

\begin{document}
\usetikzlibrary{decorations} 

\numberwithin{equation}{section}


\thispagestyle{empty}

\begin{titlepage}
\hfill MCTP-17-16
\vspace{1cm}

\begin{center}

{\Large \bf Strings in Bubbling Geometries and Dual Wilson Loop Correlators}\\

%

\vspace{1.5cm}

{\large \bf Jerem\'ias Aguilera-Damia${}^a$, Diego H. Correa${}^a$, Francesco Fucito${}^b$}\\

\vspace{.6cm}

{\large \bf Victor I. Giraldo-Rivera${}^c$, Jose F. Morales${}^b$ and Leopoldo A. Pando Zayas${}^d$}

\vspace{1cm}

{\it ${}^a$Instituto de F\'isica La Plata, CONICET, Universidad Nacional de La Plata C.C. 67, 1900}\\
{\it La Plata, Argentina}

\vspace{.4cm}

{\it ${}^b$I.N.F.N - sezione di Roma Tor Vergata
Via della Ricerca Scientifica, I-00133 Roma, Italy}

\vspace{.4cm}
{\it ${}^c$International Centre for Theoretical Sciences (ICTS-TIFR), Shivakote, Hesaraghatta Hobli,}
{\it Bengaluru 560089, India}

\vspace{.4cm}
{\it ${}^d$Michigan Center for Theoretical Physics, Randall Laboratory of Physics,}\\
{\it The University of Michigan, Ann Arbor, MI 48109-1040, USA}\\

\vspace{.4cm}
{\it ${}^d$The Abdus Salam International Centre for Theoretical Physics,
Strada Costiera 11,}\\
{\it  34014 Trieste, Italy}

\vspace{14pt}

\end{center}
\begin{abstract}
We consider a fundamental string in a bubbling geometry of arbitrary genus dual to a half-supersymmetric Wilson loop in a general large  representation $\mathbf{R}$ of the $SU(N)$ gauge group in ${\cal N}=4$ Supersymmetric Yang-Mills. We demonstrate, under some mild conditions, that the minimum value of the string classical action for a bubbling geometry of arbitrary genus precisely matches the correlator of a Wilson loop in the fundamental representation and one in a general large representation. We work out the case in which the large representation is given by a rectangular Young Tableau, corresponding to a genus one bubbling geometry, explicitly. We also present explicit results in the field theory for a correlator of two Wilson loops: a large one in an arbitrary representation  and a ``small" one in the fundamental, totally symmetric or totally antisymmetric representation.


\end{abstract}

\end{titlepage}

\setcounter{page}{1} \renewcommand{\thefootnote}{\arabic{footnote}}
\setcounter{footnote}{0}
\newpage

\tableofcontents

\section{Introduction}

In the best understood examples of the AdS/CFT correspondence, the gravity description is accomplished in terms of strings and D-branes in Anti de Sitter spaces with a constant dilaton, reflecting the conformal symmetry of the quantum field theory description. Considering heavy objects on the gravity side naturally leads to backreaction in which case the isometries of AdS are only preserved asymptotically and the dilaton is no longer constant.  In the quantum field theory description this situation corresponds, typically, to the computation of expectation values, not in the vacuum of the theory, but in some states related to operators with large quantum numbers.  This setup deviates from conformal invariance and in this manuscript we explore one of its explicit still controlled instances.

When deviating from strict AdS spaces there are not as many exact results as in conformal situations where one can explore the scenario described in the previous paragraph by comparing string theory with gauge theory results explicitly. One rare example of such exact results in non-conformal situations is the computation of the partition function, Wilson loops expectation values and correlators in ${\cal N}=2^{*}$ super Yang-Mills and its holographic dual \cite{Buchel:2013id,Bobev:2013cja,Chen-Lin:2015dfa,Chen-Lin:2015xlh,Liu:2017fiq}.

A different setup to study the AdS/CFT correspondence in non-conformal situations, which we intend to explore in this article, arises with the computation of Wilson loop correlators in cases where  one of them is taken in a large rank representation. On the gravity side, such large rank representation Wilson loops are described in terms of $\tfrac{1}{2}$-BPS backreacted spaces, with isometry group $SO(2, 1)\times SO(3)\times SO(5)$ and which present a running dilaton and fluxes turned on. The construction of these bubbling geometries (see \cite{Lin:2004nb} for bubbling geometries associated to the insertion of chiral fields) took various steps \cite{Lunin:2006xr,Yamaguchi:2006te} before culminating in \cite{D'Hoker:2007fq},
where these type IIB supergravity solutions were found in terms of two harmonic functions on a Riemann surface $\Sigma$ on whose boundary the dual Wilson loop representation data is encoded. These supergravity  solutions are highly involved and arguably represent the state-of-the-art as a far as supergravity solutions are concerned. Strings and minimal area surfaces in this kind of bubbling geometries have been studied in \cite{Benichou:2011aa,Gentle:2014lva}, in order to compute gravitational potential between open strings and to account for entanglement entropies holographically.

The expectation value of $\tfrac{1}{2}$-BPS circular Wilson loops for arbitrary representations can be computed with a Gaussian matrix model. This was first conjectured by Erickson, Semenoff and Zarembo in \cite{Erickson:2000af} and Drukker and Gross in \cite{Drukker:2000rr}, and it was finally proven by Pestun  using supersymmetric localization \cite{Pestun:2007rz}.  Remarkably, if the Wilson loop  is taken in the fundamental representation, the matrix model solution leads to an explicit expression via orthogonal polynomials which is exact in the 't Hooft coupling $\lambda$ as well as in the rank of the gauge group, $N$, \cite{Drukker:2000rr}. For higher rank representations the holographic dictionary was established in   \cite{Gomis:2006sb,Gomis:2006im}, however, with few exceptions \cite{Fiol:2013hna},  exact expressions for generic $\lambda$ and $N$ seem currently out of reach. Nevertheless, for totaly symmetric and antisymmetric representations, it is possible to obtain expressions that hold in the planar and large $\lambda$ limit \cite{Hartnoll:2006is}, that successfully match the associated D-branes on-shell actions \cite{Drukker:2005kx,Yamaguchi:2006tq}, as predicted by the AdS/CFT correspondence.
Later on, localization techniques were used for other kinds of Wilson loops of arbitrary shapes, preserving less supersymmetry  \cite{Drukker:2007dw,Drukker:2007yx,Drukker:2007qr} or to account for correlators of supersymmetric Wilson loops \cite{Giombi:2009ms,Bassetto:2009rt,Giombi:2009ds,Bassetto:2009ms,Giombi:2012ep}, but most of the explicit results have been found for the fundamental representation.

When the Wilson loop representation is even larger, for instance, when the  associated Young tableau possesses a number of order $N^2$ boxes, the dual description involves a large number of D-branes that back-react on the geometry. The corresponding matrix model can be solved with a saddle point approximation in the large-$N$ limit provided the sizes the of  Young tableau edges $\{n_i,k_i\}$ are taken to be of order $N$ \cite{Okuda:2007kh}. The eigenvalue distribution can be determined in terms of geometric data on the spectral curve which, moreover, is identified with the hyperelliptic surface characterizing the bubbling geometry as beautifully demonstrated in \cite{Okuda:2008px}.

The main purpose of this paper is to compute correlators $\langle W_\textbf{R} W_\textbf{r} \rangle $, between Wilson loops in large representations $\textbf{R}$, whose Young tableau edges $\{n_i,k_i\}$ are of order $N$, and Wilson loops in a ``small'' representation, let us say,  fundamental, completely symmetric and completely anti-symmetric. We will consider in particular the case in which both Wilson loops are defined over the coincident circle and coupled to the same scalar, so that both are invariant under the same set of symmetries and supersymmetry transformations. This allows to compute the correlator directly in the field theory using the matrix model that is obtained by supersymmetric localization. The gist of our matrix model calculation is that the ``small''  Wilson loop does not back-react on the eigenvalue distribution of the large representation Wilson loop. Thus, the correlator is eventually given by an expectation value in the eigenvalue distribution of the large representation Wilson loop.

According to the AdS/CFT correspondence, the correlator of Wilson loops of the form $\langle W_{\mathbf{R}} W_{\text{fund}}\rangle$ can be computed, in the large 't Hooft coupling $\lambda$ limit, as the on-shell action of certain strings in the bubbling geometries found in \cite{D'Hoker:2007fq}. Among the many strings that can propagate in the bubbling geometries, the ones that can be related to the particular correlator given are those invariant under the same symmetries and supersymmetries of the background. We demonstrate in this manuscript that there is precise agreement between the two sides of the correspondence.

The paper is organized as follows. In section \ref{reviewbubbling} we review the bubbling geometries dual to large representation Wilson loops and the relation between their charges and the Young tableau parameters. In section \ref{strings} we present the minimal area string configurations in generic bubbling geometries. We consider in detail the case of strings in genus one bubbling geometries, dual to a Wilson loop in a rectangular Young tableau representation, and give explicit expressions for the on-shell actions that will be later compared with matrix model results.  At the end of this section, we extend our results to general genus $g$ backgrounds. In section \ref{matrixmodel} we turn to the matrix model description of the correlator of Wilson loops. We first focus on the correlator of a Wilson loop in the fundamental representation and one in a representation given by a rectangular Young tableau, but we later consider more generic cases. We finally conclude and comment our results in section \ref{conclu}. We also include various appendices for the readers interested in further details on the results presented in the main text.


\section{ Review of bubbling geometries dual to ${1\over 2}$-BPS Wilson loops}
\label{reviewbubbling}

The general bubbling geometry background corresponds to solutions of type IIB supergravity that preserves a $SO(2,1)\times SO(3)\times SO(5)$ isometry group and 1/2 of the total supersymmetry \cite{D'Hoker:2007fq}. The resulting metric is the one associated with an $\mathbb{H}_2$, $S^2$ and $S^4$ fibration over a 2-dimensional complex Riemann surface $\Sigma$. The metric in the Einstein frame can be written as
\be
ds^2 = G^E_{MN} \,dx^M\, dx^N=f_1^2 ds^2_{\mathbb{H}_2}+f_2^2ds^2_{S^2} +f_4^2ds^2_{S^4} +d\Sigma^2\,.
\label{metric}
\ee
A quite remarkable fact about these solutions is that all the geometric functions and fluxes are completely determined by two holomorphic functions $\mathcal{A}$ and $\mathcal{B}$ defined on the Riemann surface $\Sigma$. Equivalently, the geometry can be specified in terms of four real harmonic functions defined as
\begin{align}
h_1 &=\mathcal{A}+\bar{\mathcal{A}}\,,   \quad   & \widetilde{h}_1 =\,{\rm i}\,\left(\mathcal{A}-\bar{\mathcal{A}}\right)\,, \nn\\
h_2 &=\mathcal{B}+\bar{\mathcal{B}}\,, \quad   & \widetilde{h}_2 =\,{\rm i}\,\left(\mathcal{B}-\bar{\mathcal{B}}\right)\,.
\label{h}
\end{align}

There are various ways of describing functions on a Riemann surface \cite{farkas2012riemann}.
For example, as functions in the upper half-plane with $g+1$ branch cuts satisfying appropriate boundary conditions. This formulation  usually provides a clearer scheme for describing general properties of the geometry. Alternatively functions $h_1$ and $h_2$ can be represented in terms of hyperelliptic functions of  the $2g$-periodic variables $(z,\bar{z})$ on a genus $g$ Riemann surface without boundaries. Along this article we will alternate between both descriptions and refer to the background with metric (\ref{metric}) generically as the genus $g$ solution.

Consider $\Sigma$ to be the half plane described by coordinates $(u,\bar{u})$. The main properties of an arbitrary genus $g$ solution are encoded in the boundary conditions satisfied by the harmonic functions over the real axis. More precisely, the $h_2$ function satisfies Dirichlet boundary conditions all along the boundary of $\Sigma$, whereas $h_1$ satisfies alternating Dirichlet and Neumann boundary conditions. The points where the  boundary condition changes are denoted by $\tilde{e}_a$ and determine the position of the branch cuts. A genus $g$ solution is obtained for a Riemann surface $\Sigma$ with $2g+2$ branch points on its boundary. It is customary to use conformal symmetry to bring a branch point, let us say $\tilde{e}_{2g+2}$, to minus infinity and consider the ordering $\tilde{e}_{2g+2}<\ldots<\tilde{e}_2<\tilde{e}_1$. Additionally, the remaining branch points are subjected to the constraint $\sum_{a=1}^{2g+1}  \tilde{e}_a=0$.

The general form of these functions satisfy the following equations
\be
\partial_u  h_1(u)= \frac{{\rm i}\, P(u)}{(u-u_0)^2\, s(u)}\,, \qquad \partial_u  h_2 (u)=-\frac{\rm i}{(u-u_0)^2}\,,
\label{udef}
\ee
where $u_0$ is a singular point where the geometry is asymptotically $AdS_5\times S^5$, $P(u)$ is a polynomial of degree $g+1$ with real coefficients and
\be
s(u)^ 2=(u-\tilde{e}_1) \prod_{i=1}^{g} (u-\tilde{e}_{2i})(u-\tilde{e}_{2i+1})\,.
\ee

Alternatively, making a conformal transformation one can get rid of the pole at the singular point. We will denote these coordinates as  $(v,\bar{v})$, for which a direct relation with the matrix model resolvent $w(x)$ can be established \cite{D'Hoker:2007fq,Okuda:2008px}.
\be
{\cal A}(v)= { {\rm i}\, \alpha' \over 8\, g_s} \,   \left[  2 \, v-w(v)  \right]   \,, \qquad   {\cal B}(v)=\frac{ {\rm i}\, \alpha'\, v}{4}   \,.
\label{AB}
\ee
In order to follow the same conventions as in \cite{Okuda:2008px}, we use $e_a$ to denote the branch point locations in $(v,\bar{v})$ coordinates.
Clearly, the use of $u$ or $v$-coordinates is a matter of taste  with no significant difference in the physical picture. Turning to the $(z,\bar{z})$ formulation, we can write
\be
d\Sigma^2=4\sigma^2 dzd\bar{z},
\label{rieman1}
\ee
where the radius $\sigma$ is a real function of $(z,\bar{z})$. The warping functions $f_1$, $f_2$, $f_4$, $\sigma$ and dilaton $\Phi$ are given by\footnote{Note that conventions in \cite{D'Hoker:2007fq,Benichou:2011aa} is $\phi=\Phi/2$. }
\bea
f_1^4= -4e^\Phi h_1^4\frac{W}{N_1}\,, \quad f_2^4= 4e^{-\Phi} h_2^4\frac{W}{N_2}\,, \quad f_4^4= 4e^{-\Phi} \frac{N_2}{W}\,, \quad  \sigma^8 = -\frac{WN_1N_2}{h_1^4 h_2^4}\,, \quad
e^{2\Phi} = -\frac{N_2}{N_1}\,,
\label{Eq:warps}
\eea
where
\begin{align}
N_1 &= 2\, h_1 \, h_2 | \partial h_1|^2    -h_1^2\, W\,,
\!\!\!\!\! &&    W       = \partial h_1 \,\bar \partial h_2 +  \partial h_2\, \bar \partial h_1  \,, \nn\\
 N_2 &= 2\,h_1 \,h_2\, |\partial h_2|^2     -h_2^2 \, W \, ,
\!\!\!\!\!&& \ V=  \partial h_1 \bar \partial h_2 -  \partial h_2 \bar \partial h_1\,.
\end{align}
and $\partial=\partial_z$, $\bar\partial=\partial_{\bar z}$.
Also the NS and RR fluxes can be written in the following way
\be
H_{3}=dB_{2}\,, \quad F_{3}=d C_{2}\, , \quad  F_5=dC_4+\frac18
\left(B_2\wedge F_3-C_2\wedge H_3\right),
\ee
and the corresponding potentials are
\be
 B_2= b_1\, \hat{e}_{\mathbb{H}_2}\,, \qquad  C_2= b_2\, \hat{e}_{S^2}\,, \qquad C_4=-4\,j_1\,\hat{e}_{\mathbb{H}_2} \wedge \hat{e}_{S^2}+4\, j_2\, \hat{e}_{S^4}\,,
\label{flux}
\ee
where $\hat{e}_{\mathbb{H}_2}$, $\hat{e}_{S^2}$ and $\hat{e}_{S^4}$ are the unit volume elements of $\mathbb{H}_2$, $S^2$ and $S^4$, respectively and
\begin{align}
b_1 &=-2 \, {\rm i}\, \frac{h_1^2\, h_2\, V}{N_1}-2\widetilde{h}_2-b_1^0\,,\nn\\
 b_2 &=-2\, {\rm i}\frac{h_1\,h_2^2\, V}{N_2}+2\,\widetilde{h}_1-b_2^0\,, \nn\\
j_2 &= {\rm i} \,h_1\,h_2\, \frac{V}{W}-\frac32 \left(\widetilde{h}_1\,h_2-h_1\widetilde{h}_2\right)+3\,{\rm i}\left(\mathcal{C-\bar{C}}\right)\,. \label{b}
\end{align}
 with $d \mathcal{C}=\mathcal{B}\partial\mathcal{A}-\mathcal{A}\partial\mathcal{B}$.
The integration constants $b_1^0$, $b_2^0$ are gauge redundancies  that  will be fixed later by requiring  that the two-form fluxes precisely vanishes at the $AdS_5$ singular point, {\it i.e.} $b_1(z_0)=b_2(z_0)=0$.
The function $j_1$ can be computed by using the self-duality of the RR 5-form obtaining
\begin{align}
\partial j_1 &= -{\rm i} \frac{f_1^2\,f_2^2}{f_4^4}\partial j_2 +\frac18 \left(b_1 \, \partial b_2-b_2 \, \partial b_1\right)\,.
\end{align}

\subsection{Charges and representation parameters}
To complete the description of the solution we find it convenient to go back to the $(u,\bar{u})$ formulation.  The harmonic function $h_1$ satisfies Dirichlet boundary conditions on the intervals $ (\tilde{e}_{2i+1}, \tilde{e}_{2i})$  and Neumann boundary conditions on the intervals $(\tilde{e}_{2i},\tilde{e}_{2i-1})$ for $i,j=1,\ldots, g+1$. Moreover, the $S^2$ and $S^4$ spheres shrink to zero size along Neumann and Dirichlet intervals respectively, as can be seen from the relation between the warping factors $f_i$ and the functions $h_i$ in Eq. (\ref{Eq:warps}).

The free parameters of the solutions, {\it i.e.} the positions and lengths of branch cuts  can be related to the lengths of the rows and columns of the Young Tableau associated to the representation of the dual Wilson loop. However, the precise relation is in general very involved and can be established through flux integrals over the non-trivial cycles of the geometry. We shall present here some general aspects for arbitrary genus and leave a more detailed discussion of this relation for the genus one example described in section \ref{strings}. A fairly complete treatment of this subject can be found in \cite{D'Hoker:2007fq,Benichou:2011aa} and we will mainly follow the ideas presented there.

\begin{figure}[h]
\centering
\def\svgwidth{10cm}
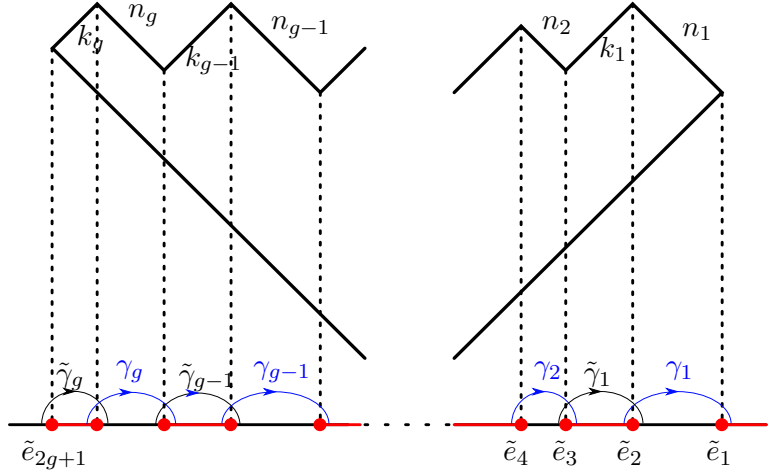
\caption{Branch cuts and generic Young tableau assigned to the dual Wilson loop. Representation parameters $\{k_j,n_i\}$ are linked to geometric parameters through flux integrals over non-trivial 3- and 5-cycles $\tilde{\gamma}_j$ and $\gamma_i$. }
\label{youngbranch}
\end{figure}

The geometric structure described so far allows to define a series of non-trivial 3- and 5-cycles encircling either Dirichlet or Neumann type intervals along the boundary of $\Sigma$\footnote{There are additional non-trivial 7-cycles given by $S^2\times \gamma_i$ and $S^4\times \tilde{\gamma}_j$ warped products which measure the fundamental string charges of the D-brane configuration\cite{Benichou:2011aa}. These charges are in turn related to the number of boxes contained in each sub-diagram of the Young tableau associated to the dual Wilson loop.   }.
Such 3- and 5-cycles have topology $S^3$ and $S^5$ respectively hence being charged under either 3- or 5-form RR fluxes. More precisely, we define the 5-cycle $\gamma_i$ as the fibration of an $S^4$ over the contour surrounding the Neumann interval $(\tilde{e}_{2i}, \tilde{e}_{2i-1})$. Analogously, the 3-cycle $\tilde{\gamma}_j$ corresponds to an $S^2$ fibration over the contour around the Dirichlet interval $(\tilde{e}_{2i+1},\tilde{e}_{2i})$. The corresponding charges can be computed by the following integrals
\begin{align}
Q^{i}_{\text{D3}}&=\oint_{\gamma_i}d C_4 \, ,\\
Q^{j}_{\text{D5}}&=\oint_{\tilde{\gamma}_j} F_3
 \end{align}
Using the Cauchy theorem and expanding the fluxes near the boundary, the integrals above can be deformed to the following integrals over the branch cuts \cite{Benichou:2011aa}:
\begin{align}
Q^{i}_{\text{D3}}&=12 {\rm i}\, \text{Vol}(S^4)\int_{\tilde{e}_{2i}}^{\tilde{e}_{2i-1}}d\mathcal{C} + \text{c.c.}\label{QD3}\, ,\\
Q^{j}_{\text{D5}}&=2 {\rm i}\, \text{Vol}(S^2)\int_{\tilde{e}_{2j+1}}^{\tilde{e}_{2j}}d\mathcal{A}+\text{c.c.}\label{QD5}\, ,
\end{align}
where
\be
d \mathcal{C}=\mathcal{B}\partial\mathcal{A}-\mathcal{A}\partial\mathcal{B}.
\ee
These integrals giving the D5 and D3 RR charges are naturally associated with the Wilson loop representation parameters (see Fig. \ref{youngbranch}) in the following way
\be
Q_{\text{D3}}^i=(4\pi^2\alpha')^2 n_i\, , \quad Q_{\text{D5}}^j=-(4\pi^2\alpha') k_j\,     \label{nkd3d5}
\ee

\section{Strings in bubbling geometries}
\label{strings}

Let us introduce a fundamental string in the bubbling geometry background just presented in the previous section and search  for minimal area solutions. Our interest in these configurations is kindled by the fact that the corresponding on-shell action can be related to the correlator of two Wilson loops, one in the fundamental representation whose dual is the fundamental string and the other in some large rank representation whose holographic dual is the background bubbling geometry itself. More precisely, in the large 't Hooft coupling limit
\be
 \langle W_{\text{fund}}\rangle_{\mathbf{R}}=\frac{\langle W_{\textbf{R}} W_{\rm fund} \rangle}{\langle W_{\textbf{R}} \rangle} \simeq \sum_{ \{ z^* \} } e^{-S_{\text{on-shell}}(z^*)}\,,
\label{correstring}
\ee

relating the correlator between the Wilson loops in the large 't Hooft coupling limit to the
gravity partition function evaluated at the points $\{ z^* \}$ of minimum action for the
fundamental string in the bubbling background. In general there will be many different classical string embeddings in a genus $g$ background, which should correspond to different specifications of the fundamental Wilson loop $W_{\rm fund}$, namely different curves and orientations in the internal space.

Since we would like to eventually   compare string theory with matrix model results, we shall focus on string configurations corresponding to fundamental Wilson loops preserving the same $SO(2,1)\times SO(3)\times SO(5)$ symmetry as the large rank representation one. This is necessary for the two Wilson loop operators to preserve the same set of supercharges. The restriction on the symmetries implies that both Wilson loops should be taken on coincident circles (with one orientation or the other)
and with same or opposite internal space orientations. Therefore, we will in turn restrict our attention to very specific dual classical string configurations.

To explicitly compare with matrix model results, we will find particular examples of these configurations and evaluate their on-shell actions. To build up our intuition we first present the general set up for the calculation and then turn to explicit examples for genus zero and  one.

\subsection{General set up}

Our aim is to solve the equations of motion derived from the Nambu-Goto action
\be
S=\frac{1}{2\pi \alpha'} \int d^2\sigma  \sqrt{ {\rm det} (G^{(S)}_{MN} \partial_\alpha X^M \partial_\beta X^N) } +\frac{1}{2\pi \alpha'} \int P\left[B_2\right]\label{nambugoto}\,,
\ee
with $G_{MN}^{(\text{S})}$ the metric in the string frame related to that one in the Einstein frame via $G^{(\text{S})}=e^{\frac{\Phi}{2}}G^{(\text{E})}$.
$P[B_2]$ is the pull-back of the NS 2-form flux over the worldsheet\footnote{Being metric independent, the coupling of the string to the $B$-field in the action remains unchanged in the new frame.}.

We consider string world sheets  extended all along the $\mathbb{H}_2$ factor parameterized by global coordinates $(\rho,\phi)$ such that $ds^2_{\mathbb{H}_2}=d\rho^2+\sinh^2\rho \,d\phi^2$ and sitting at an arbitrary point on both the $S^2$ and $S^4$.  Notice that, given this parametrization for the $\mathbb{H}_2$ factor, the corresponding string describes a circular contour on the $AdS$ boundary\footnote{Recall that, in global coordinates, the regularized $\mathbb{H}_2$ volume is finite and equals to $-2\pi$.  Should we have taken the $\mathbb{H}_2$ factor in Poincar\'e coordinates, then the regularized volume would be zero. This last parametrization is associated to a single straight Wilson line, which has trivial vacuum expectation value $\langle W\rangle=1$. }. Furthermore, we work in the formulation where $\Sigma$ is a genus $g$ Riemann surface described by coordinates $(z,\bar{z})$ which we further assume can only depend on the worldsheet coordinate $\rho$. Plugging this {\it ansatz} into Eq. \eqref{nambugoto} and using the explicit form for both the metric and the antisymmetric tensor given in Eqs. \eqref{metric}, \eqref{flux} and \eqref{b} yields
\be
S=\frac{1}{2\pi \alpha'} \int d\phi\, d\rho\, \sinh\rho \, e^{\frac{\Phi}{2}}{f_1^2}\,\sqrt{1+\frac{4\,\sigma^2 }{ f_1^2} |z'|^2} + \frac{1}{2\pi \alpha'} \int d\phi \,d\rho \, \sinh\rho \, b_1\, ,
\ee
with $z'={dz}/{d\rho}$.
The Euler-Lagrange equation becomes
\begin{align}
 \partial_z \left( e^{\Phi\over 2} f_1^2 \right)\sqrt{1+\frac{4\sigma^2 |z'|^2 }{ f_1^2}}
+  e^{\Phi\over 2} f_1^2 \partial_z  \sqrt{1+\frac{4\sigma^2 |z'|^2 }{  f_1^2}} +\partial_z b_1 =\frac{1}{\sinh\rho}\frac{d}{d\rho}\left(\frac{2e^{\frac{\Phi}{2}}\sigma^2 \bar{z}'}{\sqrt{1+\frac{4\sigma^2|z'|^2  }{  f_1^2}}}\right)\,.
\label{Eq:motion}
\end{align}

Although finding a general solution to the above equation looks like a daunting task in the general case,  there is a particularly simple solution. Indeed, if there is a point $z=z^*$ in the Riemann surface such that
\be
 \partial_z \left( e^{\Phi\over 2} \,f_1^2 \right)=\partial_z b_1=0 \label{saddle}\,,
\ee
then keeping $z=z^*$ constant, {\it i.e.} $z'=0$, gives a solution of the equations of motion. Fortunately, solutions with the aforementioned symmetry restrictions will be found  within this class. For these solutions the on-shell action reads
\be
S_{\text{on-shell}}=\left.\frac{{\rm vol}({AdS_2})}{2\pi \alpha'} \left( e^{\Phi\over 2} \, f^2_1+b_1\right)\right|_{z=z^*}
=\left.-\frac{1}{\alpha'} \left( e^{\Phi\over 2} \, f^2_1+b_1\right)\right|_{z=z^*}\,,
\label{stringaction}
\ee
where   we used the regularized volume $ {\rm vol}({AdS_2})=-2\pi$.

At this point we would like to come back to the issue of fixing the gauge ambiguity of the background fluxes. In particular, a gauge transformation of the $B$-field changes the string action by a boundary term, thus leaving
the classical configurations unaffected because the equations of motion remain invariant. However, the gauge choice does affect the evaluation of the on-shell action. As already mentioned, we fix the gauge redundancy of the $B$-field by requiring that $b_1(z_0) = 0$. This means that the $B$-field vanishes at the singular point where the background is asymptotically $AdS_5\times S^5$, thus being identified with the dual CFT vacuum. Otherwise, if $b_1(z_0)$ were non-vanishing,
a non-trivial source should be turned on at the boundary CFT that would take us away from the vacuum.

In the following subsections we will find classical solutions to the Euler-Lagrange equations and evaluate the on-shell action for strings in genus zero and genus one  supergravity backgrounds.

\subsection{Strings in genus zero background}

To familiarize the reader with the details of the presentation of the solution we review the computation of a minimal string area on $AdS_5\times S^5$, which corresponds to the genus zero background geometry. Despite being a well known result, a reformulation of this problem in the geometrical language just presented in the previous section would introduce some hints about the manipulations that we will perform in the genus one case.

The $AdS_5\times S^5$ solution in the $(v,\bar{v})$ formulation is obtained by taking
\be
{\cal A} =- \frac{ \alpha' }{4 \, g_s }\, \sqrt{\lambda-  v^2   } \, \qquad~~~~~~~~~
 {\cal B} = {\rm i} \frac{ \alpha'\, v }{4} .
\label{Eq:h-genus0}
\ee
with $\alpha'$, $\lambda$ and $g_s$ related to the radiue $L$, the RR flux $N$ and the dilaton $\Phi_0$ of the $AdS_5\times S^5$ solution via\footnote{ $L^4$ is proportional to $N$ in the Einstein frame and to $\lambda$ in the string frame.}
\be
L^4=4\pi N \alpha'^2\quad , \quad  e^{\Phi_0}=g_s\quad , \quad  \lambda=4\, \pi\, g_s\, N
\ee
More precisely, plugging (\ref{Eq:h-genus0}) one finds the dilaton and warping factors
 \bea
 f_1^2-f_2^2=L^2,\quad\quad~~~~~~~ \sigma^2={ L^2 \over 4 |1- { v^2 \over \lambda} | },   \quad\quad~~~~~~ e^{\Phi}=e^{\Phi_0},  \label{39}
 \eea
 The gauge fixed $B$ field is vanishing.
Note that $h_1={\cal A}+{\cal \bar A}$ satisfies Neumann boundary conditions along the real segment $(-\sqrt{\lambda},\sqrt{\lambda})$ and Dirichlet along the remaining segments of the real axis.  Moreover, given (\ref{39}), we note that  $f_1$ becomes constant wherever $f_2$ vanishes, namely for $v^*\in[-\sqrt{\lambda},\sqrt{\lambda}]$. Therefore, any point lying on this segment corresponds to a solution of the equations of motion. Furthermore, all these solutions lead to the same on-shell action
\be
S_{\text{on-shell}}=-\frac{ e^{\Phi_0 \over 2} \, f^2_1(v^*)}{\alpha'}=-\frac{ e^{\Phi_0 \over 2}  \,  L^2 }{\alpha'}=-\sqrt{\lambda}\,,
\ee
 From the foliation of the solution it should be clear that the Riemann surface provides the radial coordinate for $AdS_5$ to be written as a foliation of $AdS_2 \times S^2$ and the angular coordinate to write $S^5$ as a foliation of $S^4$. This becomes evident if we perform the following change of variables
\be
 v = \sqrt{\lambda}\,  \cosh(\eta-{\rm i \,}  \theta)\, , \quad 0\leq \eta<\infty\, , \quad  0\leq\theta\leq\pi,
\ee
under which the metric takes the familiar form
\be
ds^2= L^2\,\left(d\eta^2 + \cosh^2\eta\, ds^2_{\mathbb{H}_2}+\sinh^2\eta\, ds^2_{S^2}+d\theta^2+\sin^2\theta\, ds^2_{S^4}\right)\,.
\ee
On the other hand, the solution segment $v^*\in   [-\sqrt{\lambda},\sqrt{\lambda} ]$ gets mapped to the segment $\eta=0\,,\, 0\leq\theta\leq\pi$ thus making manifest that different choices of $v^*$ correspond to different polar angles on the $S^5$. In particular the branch points $v^*=\pm \sqrt{\lambda} $ corresponds to the north and south poles of $S^5$ and solutions placed at these points will be dual to configurations associated to Wilson loops coupled with
opposite orientation in the six-dimensional internal space.

\subsection{Strings in genus one backgrounds}

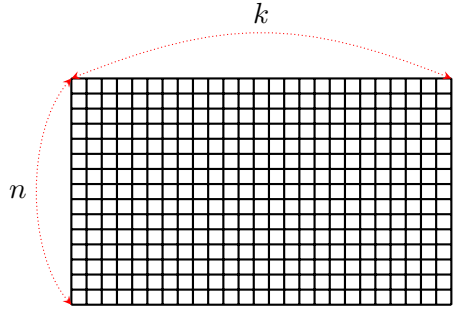
\begin{figure}[h!]
\beginpgfgraphicnamed{Rec-Rep}
\begin{center}
\begin{tikzpicture}[>=stealth]
\draw[thick] (0,0) grid  [xstep=.2,ystep=.2]  (5,3);
\draw[densely dotted,<->,red] (0,3) .. controls (2,3.8) and (3,3.8) .. (5,3) node[above,pos=.5,black]{$k$};
\draw[densely dotted,<->,red] (0,3) .. controls (-0.6,2.3) and (-0.6,0.7) .. (0,0) node[left,pos=.5,black]{$n$};
\end{tikzpicture}
\end{center}
\endpgfgraphicnamed
\caption{Number of rows and columns in the tableau are related to the charges $Q^1_{\text{D3}}$ and $Q^1_{\text{D5}}$.}
\label{fig:RecRep}
\end{figure}

In this section we will consider genus one backgrounds since they can be explicitly realized in terms of Weierstrass elliptic functions \cite{D'Hoker:2007fq}. These geometries arise due to the backreaction of a Wilson loop in a representaion given by a rectangular Young tableau with $n_1=n$ rows and $k_1=k$ columns, see Fig.\ref{fig:RecRep}. In this case, the most convenient approach corresponds to taking $\Sigma$ as a torus described by coordinates $(z,\bar{z})$ with periods $2 \omega_1$ and  $2 \omega_3$.  The Weierstrass elliptic functions provide the mapping between the torus and the half complex plane. In particular, taking $z_0=1$, the holomorphic functions take the form
\begin{align}
{\cal A}&= {\rm i}\, \kappa_1 \left(\zeta(z-1)+\zeta(z+1)-2\frac{\zeta(\omega_3)}{\omega_3}z  \right)\,,
\nn
\\
{\cal B} &= {\rm i}\, \kappa_2 \left(\zeta(z-1)-\zeta(z+1) \right)\,,
\label{h2}
\end{align}
where $\zeta$ denotes the Weierstrass $\zeta$-function, a primitive of the Weierstrass $\wp$-function
\be
\wp(z)=-\zeta'(z)\,,
\ee
  satisfying the condition $\lim_{z\to 0} (\zeta(z)-1/z)=0$.  The functions $\zeta(z)$ and $\wp(z)$ depend implicitly on two numbers $g_2,g_3$ (or equivalently $\tilde{e}_1,\tilde{e}_2$ ) specifying the periods of the torus. More precisely,  $\wp(z)$  can be defined as the solution of the differential equation
  \be
  \left[ \wp'(z) \right]^2=4  \left[ \wp(z) \right]^3-g_2\,  \wp(z)  -g_3=4  \left[ \wp(z)-\tilde{e}_1 \right] \left[ \wp(z)-\tilde{e}_2 \right] \left[ \wp(z)-\tilde{e}_3 \right] \label{diff}\,,
  \ee
  with $\tilde{e}_1+\tilde{e}_2+\tilde{e}_3=0$ and
  \be
g_2=2\left(\tilde{e}_1^2+\tilde{e}_2^2+\tilde{e}_3^2\right)\,, \qquad  g_3= 4 \tilde{e}_1 \tilde{e}_2 \tilde{e}_3\,.
\ee
At the half periods, $\omega_i$, one finds $\wp(\omega_i)=e_i$ and  $\wp '(\omega_i)=0$,  so Eq. (\ref{diff}) is verified. Given the branch points $\tilde{e}_1,\tilde{e}_2$ one can compute the periods  $2\omega_1$ and $2\omega_3$ using the standard elliptic formulas
\be
\omega_1= \frac{ K\left(\tfrac{\tilde{e}_2-\tilde{e}_3}{\tilde{e}_1-\tilde{e}_3}\right)}{\sqrt{\tilde{e}_1-\tilde{e}_3}} \,, \qquad
\omega_3= {\rm i}\,\frac{K\left(\tfrac{\tilde{e}_1-\tilde{e}_2}{\tilde{e}_1-\tilde{e}_3}\right)}{\sqrt{\tilde{e}_1-\tilde{e}_3}} \,,\qquad \omega_2=\omega_1+\omega_3\,,
\label{halfperiods}
\ee
where $K$ is the complete elliptic integral of the first kind. Finally $\kappa_1$ and $\kappa_2$ are determined by requiring that the geometry reduces asymptotically to $AdS_5\times S^5$ when $z\to z_0=1$.
Near this point one finds
\begin{align}
{\cal A}& \underset{z\to 1}{\approx}    {\rm i}\, \kappa_1 \left[ {1\over (z-1)} +\zeta(2)-2\frac{\zeta(\omega_3)}{\omega_3} - \left(\wp(2)+2 \, {\zeta(\omega_3)\over \omega_3} \right)(z-1)  -{ \wp '(2) \over 2} (z-1)^2+\ldots \right]\,,
\nn
\\
{\cal B} & \underset{z\to 1}{\approx}    {\rm i}\, \kappa_2 \left( {1\over (z-1)} -\zeta(2) +\wp(2)(z-1)  +{ \wp '(2) \over 2} (z-1)^2 +\ldots \right)\,.
\label{nearz1}
\end{align}
Comparing with Eq. (\ref{Eq:h-genus0}), one finds that the match requires
  \begin{align}
\kappa_1&= \frac{L^2}{8} e^{-\frac{\Phi_0}{2}}
\left(\wp(2)+\frac{\zeta(\omega_3)}{\omega_3}\right)^{-\frac12}, \\
\kappa_2&=\frac{L^2}{8}e^{\frac{\Phi_0}{2}}
\left(\wp(2)+\frac{\zeta(\omega_3)}{\omega_3}\right)^{-\frac12}\,.
\end{align}
 Moreover, requiring that $b_1=0$ at $z=1$ one finds
 \be
b_1^0 = 2\kappa_2\left(\frac{\wp'(2)}{\wp(2)+\frac{\zeta(\omega_3)}{\omega_3}}-2\zeta(2)\right)\,.
\label{gaugefix}
\ee

\begin{figure}[h]
\centering
\def\svgwidth{12cm}
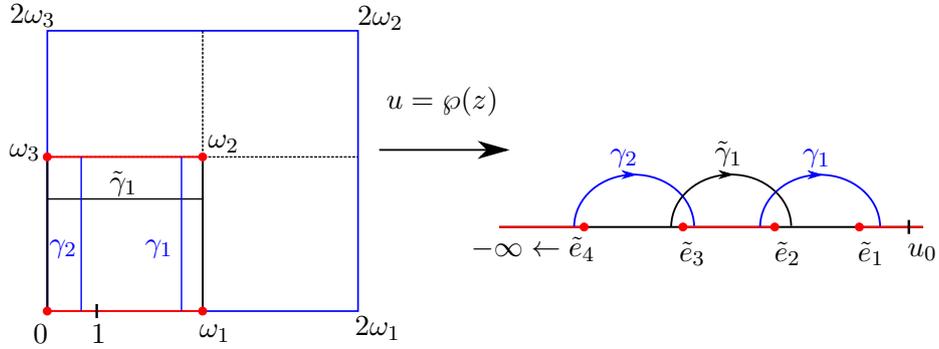
\caption{Mapping from the torus to the half-plane. The boundary of the fundamental domain of the Weierstrass elliptic functions delimited by $\{0,\omega_1,\omega_2,\omega_3\}$ gets mapped to the boundary at the real axis ($\wp(\omega_i)=\tilde{e}_i$).}
\label{torus}
\end{figure}

\vspace{0.5cm}
The number of rows and columns in a rectangular Young tableau are directly related to the charges $Q^1_{\text{D3}}$ and $Q^1_{\text{D5}}$ of the supergravity solution, given by the expressions \eqref{QD3} and \eqref{QD5} respectively, while the rank $N$ of the gauge group is related to  $Q^0_{\text{D3}}=Q^2_{\text{D3}}+Q_{\text{D3}}^1$. Indeed, for the genus one case there are two non-trivial 5-cycles $\gamma_1$ and $\gamma_2$ and one non-trivial 3-cycle $\tilde{\gamma}_1$ (see Figure \ref{torus}), these charges have been computed explicitly \cite{Benichou:2011aa} obtaining\footnote{Here we used formula D.8
of  \cite{Benichou:2011aa} and the identity $ \wp(2) = \tfrac{1}{4}\left(\frac{\wp ''(1)}{\wp '(1)}\right)^2 -2 \wp(1) $. }
\bea
N -n\!&=&\! \frac{Q^2_{\text{D3}}}{(4\pi^2 {\alpha'})^2}\,, \nn\\
n \!&=&\! \frac{Q^1_{\text{D3}}}{(4\pi^2 {\alpha'})^2} = \frac{N \omega_3}{2\pi \, {\rm i}\,}
\left(4\left(\zeta(1)-\tfrac{\zeta(\omega_3)}{\omega_3}\right)+\frac{\left(\wp(1)+\tfrac{\zeta(\omega_3)}{\omega_3}\right)\wp''(1)-\wp'(1)^2}
{\left(\wp(2)+\tfrac{\zeta(\omega_3)}{\omega_3}\right)\wp'(1)}\right)\,,
 \nn\\
k \!&=&\! -\frac{Q^1_{\text{D5}}}{4\pi^2 {\alpha'}} = \frac{\sqrt{\pi}\, {\rm i}\,}{\omega_3}\sqrt{\frac{N}{g_s}} \left(\wp(2)+\tfrac{\zeta(\omega_3)}{\omega_3}\right)^{-1/2}\,.
 \label{N5}
\eea

In what follows let us find the solutions $z=z^*$ of  Eq. (\ref{Eq:motion}) for this particular case. Recall that we are interested in string configurations preserving the same $SO(2,1)\times SO(3)\times SO(5)$ symmetry as the background. It turns out that the only points on the Riemann surface consistent with this condition are those where both the $S^2$ and the $S^4$ shrink to zero size, which corresponds precisely to the branch points where the warping factors $f_2$ and $f_4$ vanish.

In order to show that they actually satisfy Eq. (\ref{saddle}) we consider the expansions of the holomorphic functions ${\cal A}$ and ${\cal B}$ around the four branch points located at $z=\omega_a$, $a=0,1,2,3$, with  $\omega_0 = 0$. Given the periodic property of the elliptic functions $\zeta(z+2\omega_i)=\zeta(z)+2\zeta(\omega_i)$,  formulas (\ref{h2}) drastically simplify  to
\bea
{\cal A} (z) & \underset{z\to \omega_a}{\approx} & c^A_0(\omega_a)+c_1(\omega_a) (z- \omega_a) +c_3(\omega_a) (z-\omega_a)^3 +\ldots  \nn\\
{\cal B} (z) & \underset{z\to \omega_a}{\approx} & c^B_0(\omega_a)+c_2(\omega_a) (z- \omega_a)^2 +c_4(\omega_a) (z-\omega_a)^4  +\ldots   \label{nearbranch}
\eea
with\footnote{We recall that $\wp''(z)=6\, \wp(z)^2-g_2/2$.}
\bea
c_1(\omega_a) &=& 2 \, {\rm i} \, \kappa_1 {\zeta(\omega_3) \over \omega_3 }\,,  \qquad   ~~~~c_3(\omega_a) = - \, \frac{ {\rm i} \, \kappa_1 }{ 3}  \wp ''(1+\omega_a)\,,    \nn\\
c_2(\omega_a) &=&  {\rm i} \, \kappa_2 \, \wp '(1+\omega_a)\,,    \qquad  c_4(\omega_a) = {\rm i} \, \kappa_2 \, \wp(1+\omega_a)   \wp ' (1+\omega_a)\,, \nn\\
c_0^A(\omega_a) &=&  -2 \, {\rm i} \, \kappa_1 \left( {\zeta(\omega_3) \over \omega_3 }\, \omega_a +\zeta(1-\omega_a)- \zeta(1+\omega_a) \right)\,, \nn\\
   c_0^B &=& - \, {\rm i} \, \kappa_2\,
\left( \zeta(1+\omega_a)   +\zeta(1-\omega_a)  \right)\,.
\eea
Plugging the expansions (\ref{nearbranch}) into the background fields (\ref{Eq:warps}) and (\ref{b}) we find
\bea
e^{\Phi\over 2} \, f_1^2 (z) & \underset{z\to \omega_a}{\approx} & \left|{ 2 \, {\rm i}\, c_1\,c_2^2  \over 2 \,c_2\, c_3-c_1\, c_4}\right| + {\mathcal O}\left[ (z-\omega_a)^2 \right],\nn\\
b_1(z) & \underset{z\to \omega_a}{\approx} & { 2 \, {\rm i}\, c_1\,c_2^2  \over 2 \,c_2\, c_3-c_1\, c_4} -b_1^0-2\, {\rm i}\, c_0^{B}  + {\cal O}\left[ (z-\omega_a)^2 \right]\ ,
\eea
showing that $z=\omega_a$ solves Eq. (\ref{saddle}). Moreover, the on-shell action is\footnote{One may use
$ \zeta\left(1+\omega_i\right)= \zeta
   \left(1-\omega_i\right)+2 \zeta(\omega_i)$. }
\begin{align}
& S_{\rm on-shell}(\omega_a)
 = -\frac{1}{ \alpha'} \,\frac{L^2 \sqrt{g_s} } {
4 \sqrt{\wp(2)+\frac{\zeta \left(\omega_3\right)}{\omega_3}}}    \left(  2 \zeta(2) - 2\left[ \zeta\left(1+\omega_a\right)+ \zeta
   \left(1-\omega_a\right) \right] -\frac{ \wp '(2)}{\wp(2) +{ \zeta \left(\omega_3\right)\over \omega_3} }  \right. \nn\\
   &
    +\left| \frac{3\, \wp '\left(1+\omega_a\right)\,  \left( \wp\left(1+\omega_a\right)+{\zeta
   \left(\omega_3\right) \over \omega_3} \right) }{  \wp ''\left(1+\omega_a\right)-3  \wp\left(1+\omega_a\right) \left(   \wp\left(1+\omega_a\right) +   \zeta
   {\left(\omega_3\right)\over \omega_3}  \right) }\right|
     -  \left.\frac{3\, \wp '\left(1+\omega_a\right)\,  \left( \wp\left(1+\omega_a\right)+{\zeta
   \left(\omega_3\right) \over \omega_3} \right) }{  \wp ''\left(1+\omega_a\right)-3  \wp\left(1+\omega_a\right) \left(   \wp\left(1+\omega_a\right) +   \zeta
   {\left(\omega_3\right)\over \omega_3}  \right) }  \right)  \label{action}
\end{align}

The string configurations we have found for the genus one case, and eventually their on-shell actions \eqref{action}  are written  as functions of the branch point positions $\tilde{e}_i$ . In order to make a comparison with the gauge field theory results it is necessary to express them in terms of the numbers of rows and columns $n$ and $k$ of the corresponding Young tableau.  To do this we have to invert  (\ref{N5}) to give the branch points $\tilde{e}_i$ and the half-periods $\omega_i$ in terms of $n$ and $k$. Although, the relation between the two sets of variables is pretty involved for generic values of $n$ and $k$, here, we are interested  in the precise regime, for which $n$ is order $N$ and $k$ is order $N$ or larger.

Accessing the regime of interest requires to take $\omega_3\to 0$ and $\omega_1$ to approach 2. In order to implement this limit, it is convenient to introduce
\be
\omega_1 = 2 -\frac{x}{\Lambda}\,,\qquad \omega_3 = \frac{\, {\rm i}\,\pi}{2\Lambda}\,,
\ee
and consider that $\Lambda$ is large and $x$ finite.  Inverting the formulas for the periods in the limit, one finds
\begin{align}
\tilde{e}_1 &= {\Lambda^2 \over 3} \left(1+24 \, e^{2x-\Lambda} +24 \, e^{4x-2\Lambda}+ {\cal O}(e^{6x-3\Lambda})   \right)\,,  \nn\\
\tilde{e}_2 &= {\Lambda^2 \over 3} \left(1-24 \, e^{2x-\Lambda} +24 \, e^{4x-2\Lambda}+ {\cal O}(e^{6x-3\Lambda})   \right)\,,
\end{align}
while the Weierstrass elliptic zeta function can be expressed as\footnote{Following sub-leading orders would not influence the on-shell evaluation of the action in the regime considered.}
\begin{align}
 \zeta(z)\simeq &
-\frac{\Lambda^2 z}{3}\left(1-\frac{3}{\Lambda z} \coth(\Lambda z)\right)
+8 \,\Lambda^2\, z\, e^{4 \,x-8\Lambda}\left(1 -\frac{ \sinh\left(2\Lambda z\right)}{2\, \Lambda \, z } \right)+{\cal O}(e^{6x-3\Lambda})\,,
\label{zetalim}
\end{align}
 and $\wp(z)=-\zeta '(z)$. In this limit the charges (\ref{N5}) adopt the form
%
\begin{align}
n = &\,\frac{e^{4x}}{1+e^{4x}} N \,,
\label{nlim}
\\
k = &\, \frac{2 e^{2\Lambda}}{\sqrt{\lambda}\sqrt{1+e^{4x}}} N\,.
\label{klim}
\end{align}
Similarly, if we use the expansions   (\ref{zetalim}), for the on-shell actions (\ref{action})   we find
\begin{align}
S_{\rm on-shell}(0)=S_{\rm on-shell}(\omega_3)= &\, -\frac{\sqrt{\lambda}}{\sqrt{1+e^{4x}}} + \frac{\sqrt{\lambda} e^{2\Lambda} e^{4x}}{2(1+e^{4x})^{3/2}}\,,
\nn\\
S_{\rm on-shell}(\omega_1)= S_{\rm on-shell}(\omega_2)=& \, - \frac{\sqrt{\lambda}e^{2x}}{\sqrt{1+e^{4x}}}-\frac{\sqrt{\lambda} e^{2\Lambda}}{2(1+e^{4x})^{3/2}}\,,
\end{align}
which can be put in terms of the number of rows and columns using (\ref{nlim}) and (\ref{klim})
\begin{align}
S_{\rm on-shell}(0)=S_{\rm on-shell}(\omega_3) = & -\sqrt{\lambda\left(1-\tfrac{n}{N}\right)}+\frac{k n\lambda}{4N^2}\label{sw03}\,,  \\
S_{\rm on-shell}(\omega_1)= S_{\rm on-shell}(\omega_2)=& -\sqrt{\lambda \tfrac{n}{N}}-\frac{k(N-n)\lambda}{4N^2}\label{sw14}\,.
\end{align}

We notice that the pair of solutions with $z^*=0,\omega_3$  or $z^*=\omega_1,\omega_2$ share the same on-shell action. They can be distinguished by the position of the fundamental string on $\Sigma$ and we would like to identify which correlators of Wilson loops can be related with each of them, according to the AdS/CFT correspondence. Because fundamental strings at  any of the four branch points  correspond to $SO(2,1)\times SO(3)\times SO(5)$  symmetric configurations,  they should correspond to correlators of Wilson loops on the same circle with either the same or the opposite internal space orientations. In the remaining of this section we will argue that the contributions of the saddle points $z^*=0,\omega_1$ has to be taken into account altogether for a given orientation of the fundamental string, and $z^*= \omega_2,\omega_3$ for the opposite one.

By considering an $AdS_5\times S^5$ limit of the bubbling geometry, it is possible to argue that strings at  $z^*=0$ and $z^*=\omega_3$ are the
dual description of correlators in which the fundamental Wilson loops have opposite internal space orientations. More precisely, we consider
the large $\omega_1$ limit, which corresponds to the collapse of one of the branch cuts (namely $\tilde{e}_2\to \tilde{e}_1$). In this limit, when the usual $AdS_5\times S^5$ background is restored (see Appendix \ref{AppProbe}), $z^*=0$ and $z^*=\omega_3$ become the antipodal points on the $S^5$, and strings located there
correspond to fundamental Wilson loops which couple to the scalars with opposite orientation in the internal space. Therefore, for the correlator of a back-reacting Wilson loop with a fundamental one with the same internal space orientation, either $z^*=0$ or $z^*=\omega_3$ has to be considered but not both.

The existence of four saddle point solutions is a non-trivial consequence of the genus one geometry. We will argue that for the dual one type of correlator (same or opposite internal space orientation) $z^*=\omega_1$ has to be taken into account altogether with $z^*=0$, while $z^*=\omega_3$ has to be taken into account altogether with $z^*=\omega_2$. This is related to the non-trivial topology of the target space. In particular, the definition domain of the generating functions is two-sheeted and then we need a two-fold boundary condition in order to have a well defined variational problem. Evidence that $z^*=0$ and $z^*=\omega_1$ corresponds to the same correlator in the dual CFT comes from the fact that $z^*=0$ and $z^*=\omega_1$ configurations are related by a large gauge transformation. If we consider for instance the transformation $z\to z+\omega_i$, the holomorphic functions $\mathcal{A}$ and $\mathcal{B}$ change as
\begin{align}
\mathcal{A}(z,z_0) \to &\ \mathcal{A}(z,z_0+\omega_i)+ \alpha_i \,, \qquad \alpha_1=\alpha_2=\, {\rm i}\, \frac{\pi\kappa_1}{|\omega_3|}\,, \qquad \alpha_3=0\,,\nn
\\
\mathcal{B}(z,z_0)\to &\ \mathcal{B}(z,z_0+\omega_i)+ \beta_i \,, \qquad \beta_i =\, {\rm i}\, 2\kappa_2\zeta(\omega_i)\,,
\end{align}
where we slightly changed the notation to make the position of the singular point manifest. The singular point can be shifted by a conformal transformation of the target space and, since  $\zeta(\omega_1)$ is real, the configurations at $\omega_0=0$ and $\omega_1$ are related by an imaginary shift of the holomorphic functions. Imaginary shifts on the holomorphic functions are related to large gauge transformations of the background fluxes which induce redefinitions of the charges, since they are fluxes integrals over non-trivial cycles. The relation of these gauge transformations to the Hanany-Witten  effect is discussed in \cite{Benichou:2011aa}. Since invariance under this kind of gauge transformations is expected, both configurations $z^*=0$ and $z^*=\omega_1$ should contribute to the saddle point dual to a given Wilson loops correlator. An analogous relation is found for $\omega_2$ and $\omega_3$.

\begin{figure}[H]
\centering
\def\svgwidth{13.8cm}
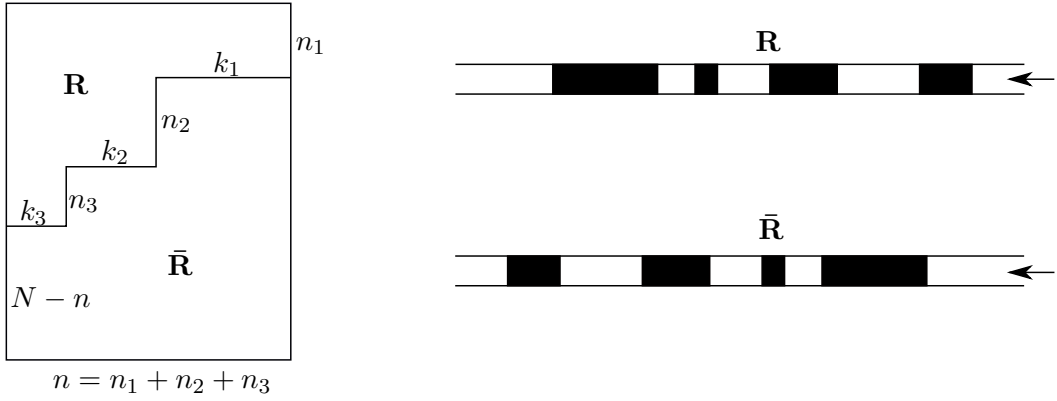
\caption{ Young tableaux for $\mathbf{R}$ and $\bar{\mathbf{R}}$ and associated Maya diagrams.}
\label{maya}
\end{figure}

\begin{figure}[H]
\centering
\def\svgwidth{15cm}
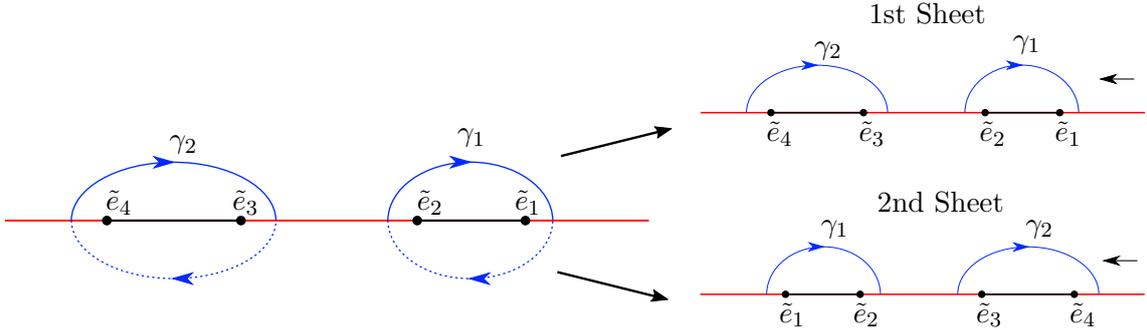
\caption{Left: Red lines denote the branch cuts and dotted blue lines indicate that cycles are closing on the second sheet of the Riemann surface. Right: Branch points and cycles interchange roles when viewing from one sheet or the other.}
\label{riemann}
\end{figure}

This gauge transformation of the background can be associated to a symmetry already present in the dual gauge theory. For a generic Wilson loop representation $\mathbf{R}$, this symmetry is the invariance under the change of $\mathbf{R}$ by its complex conjugate $\bar{\mathbf{R}}$. The conjugate representation is obtained by inverting the Maya diagram assigned to a given tableau \cite{Yamaguchi:2006te,Okuda:2007kh} (see Figure \ref{maya}). Black segments in the Maya diagram are a direct representation of the cuts of the density of eigenvalues $\rho(x)$ in the associated matrix model that will be encountered in next section.
\vspace{0.5cm}
In the gravity description, this conjugation symmetry can be interpreted as viewing the geometry from either one or the other Riemann sheet (see fig. \ref{riemann}) and the roles played by branch point $\tilde{e}_4=-\infty$ ($z=0$) and $\tilde{e}_1$ ($z=\omega_1$) are exchanged; the same occurs with the roles played by $\tilde{e}_2$ ($z=\omega_2$) and $\tilde{e}_3$ ($z=\omega_3$). Additionally, the non-trivial cycles get interchanged, giving rise to the usual $n\to N-n$ transformation.

Collecting the two contributions together and defining $\nu=\frac{n}{N}$ we can write the final AdS/CFT result for the correlator
\be
\label{eq:HolCorr}
\langle W_{\rm fund} \rangle_{\mathbf{R}} \approx e^{\sqrt{\lambda(1-\nu)}-\frac{k\nu \lambda}{4N}} +  e^{\sqrt{\lambda\, \nu}+\frac{k(1-\nu)\lambda}{4N}}
\ee
As a final remark, we notice that the result is invariant under $n\to N-n$ when also taking $k\to-k$, suggesting that the conjugation of the representation  is related
to a different choice of orientations of the brane system.

So far, as it has been stressed before, the string configurations we have found are the dual description of correlators between two Wilson loops defined along  the same circular contour with either the same or the opposite orientations in the internal space. However, this does not exhaust all the possible configurations consistent with the symmetry $SO(2,1)\times SO(3)\times SO(5)$. Indeed, we should allow for the possibility of correlators between two Wilson loops defined along  circular contours with opposite  space-time orientations  with either the same or the opposite internal space orientations.

The dynamics of a string dual to a Wilson loop with opposite space-time orientations is governed by a similar Nambu-Goto action, but with a sign changed in front of the  $B$-field term. Interestingly, the configurations at the points  $z=\omega_a$ also satisfy the Euler-Lagrange of this alternative problem. The on-shell actions for these strings with opposite  space-time orientations are
\begin{align}
\tilde S_{\rm on-shell}(0) = \tilde S_{\rm on-shell}(\omega_3) = & -\sqrt{\lambda\left(1-\tfrac{n}{N}\right)}-\frac{k n\lambda}{4N^2}\label{stw03}\,,  \\
\tilde S_{\rm on-shell}(\omega_1)= \tilde S_{\rm on-shell}(\omega_2)=& -\sqrt{\lambda \tfrac{n}{N}}+\frac{k(N-n)\lambda}{4N^2}\label{stw14}\,.
\end{align}
Reasoning as before, one can conclude that $z^*=0$ and $z^*=\omega_1$ or $z^*=\omega_2$ and $z^*=\omega_3$ contribute to this other type of correlators,
depending on the relative internal space orientation. Thus, the AdS/CFT result for this other type of correlators is
\be
\label{eq:HolCorr2}
\langle \, \widetilde W_{\rm fund} \,  \rangle_{\mathbf{R}}\approx e^{\sqrt{\lambda(1-\nu)}+\frac{k\nu \lambda}{4N}} +  e^{\sqrt{\lambda\,\nu}-\frac{k(1-\nu)\lambda}{4N}}.
\ee

We will find in the next section that the matrix model computation matches the result above, giving an indirect support to our interpretation. In appendix \ref{susywl} we study the supersymmetric properties of this configuration of Wilson loops from the field theory side.

\subsection{Strings in genus $g$ backgrounds}

Finally, we consider a fundamental string in a general genus $g$ background.  We work in the half-plane formulation, where the supergravity solution is specified by a single holomorphic function $w(v)$  in the upper half-plane with  $g+1$ cuts along the real line.  This function can be identified with the resolvent of the dual matrix model description \cite{Okuda:2008px}. We will first prove that, given a genus $g$ background geometry,  fundamental strings sitting at any of the $2g+2$  branch points $e_{a}$ give rise to solutions of the Euler-Lagrange equations and then we will evaluate the action of the fundamental string at these points.

In the $v$-plane, the functions ${\cal A}$ and ${\cal B}$ are given by
\be
{\cal A}(v)= { {\rm i}\, \alpha' \over 8\, g_s} \,   \left[  2 \, v-w(v)  \right]   \,, \qquad   {\cal B}(v)=\frac{ {\rm i}\, \alpha'\, v}{4}   \,.
\label{generalg}
\ee
In these coordinates the $AdS_5\times S^5$ asymptotic region is approached  as $v\to\infty$. The asymptotic behavior
of the holomorphic function $w(v)$ is given by
\be
w(v)=\frac{\lambda}{v}+\frac{\lambda w_1}{v^2}+\mathcal{O}(v^{-3})\,.
\label{decay}
\ee
Plugging \eqref{generalg} and \eqref{decay} into the gravity solution one finds that the potential $b_1$ vanishes for $v\to\infty$ provided
$b_1^0 = \alpha' w_1$.

Let us now consider the string action in the vicinity of the branch points $e_a$. Expansions of $h_1$ and $h_2$ near the real line have been performed in \cite{D'Hoker:2007fq}. If we write $v=x+{\rm i}\,y$ and expand all functions near the boundary $y\approx 0$, we get
\begin{align}
h_1 &={\cal A}+{\cal \bar A}= a_0(x) +a_1(x)y +  a_2(x)y^2 +a_3(x)y^3 + \mathcal{O}(y^4)\,,\nn\\
h_2 &={\cal B}+{\cal \bar B}=   -\alpha' \,{y \over 2}  \label{hdir20}\,.
\end{align}
The coefficient $a_{2k}$ and $a_{2k+1}$ are completely determined in terms of $a_0$ and $a_1$ respectively by means of the harmonic equation $(\partial_x^2+\partial_y^2)h_1=0$. In particular
\be
a_2(x)=-\frac12 a_0''(x)\, , \quad a_3(x)=-\frac16 a_1''(x)\,,
\label{harmonicsol0}
\ee
and so on. Moreover, along the real line, $h_1$ satisfies either Neumann or Dirichlet boundary conditions and therefore either $a_0(x)$ or $a_1(x)$  vanish along the real line. So one can write
\be
h_1 (x+{\rm i} y) =\left\{
\begin{array}{ccc}\label{hnd}
 a_0(x)  + a_2(x)y^2 + \ldots  &  N:&    x\in (e_{2i},e_{2i-1})  \\
 a_1(x)y  +a_3(x)y^3 +\ldots  &  D: &    x\in  (e_{2j+1},e_{2j})   \\
\end{array}
\right.
\ee
For example, approaching the real line along an interval with Neumann boundary conditions, using (\ref{hdir20}-\ref{hnd}),   we obtain the expansions
\begin{align}
W&=\alpha' \,{a_0''(x)\, y\over 4} +\mathcal{O}(y^3)\,,  \quad~ ~~~~~~~~~~~~~~~~~~~~~~~~~~~~~~~~~~~ V=-\alpha' {i a_0'(x)\over 4}+\mathcal{O}(y^2)\,,\\
N_1 &=-\alpha' \, { a_0(x)\,  \over 4} \left[ a_0'(x)^2+a_0(x)\,a_0''(x)\right]\, y + \mathcal{O}(y^3)\,,\quad  \quad
N_2 = -\frac{(\alpha' )^3}{16}a_0(x) y + \mathcal{O}(y^3)\,.
\end{align}
leading to
\begin{align}
e^{\Phi\over 2} \, f_1^2 =&\ \alpha'  \left| { \sqrt{ a_0(x)^3\, a_0''(x) } \over  a_0'(x)^2+a_0(x)\, a_0''(x) } \right| +\mathcal{O}(y^2)\,,
\nn\\
b_1 =&\    \alpha' x-\alpha'  { a_0(x)\, a_0'(x)  \over  a_0'(x)^2+a_0(x)\, a_0''(x) } - b_1^0  +\mathcal{O}(y^2)\,,
\label{actionbranch}
\end{align}
At the branch points, $h_1$ satisfies both Neumann and Dirichlet boundary conditions and therefore we impose $a_0$ to vanish there. Moreover, ${\cal A}$ has to develop a branch cut discontinuity at those points. Taking
\be
a_0(x) =  (x-e_a)^{1\over 2} \left[ C_{a,0}+C_{a,1}\,(x-e_a)+C_{a,2}\,(x-e_a)^2+ {\cal O}(x-e_a)^3  \right]\,,
\label{a0exp}
\ee
where $C_{a,i}$ are numerical coefficients.
Expanding (\ref{actionbranch}) around these points we find
\begin{align}
e^{\Phi\over 2} \, f_1^2  = &\ \alpha'    \left| {C_{a,0}\over 4\, C_{a,1}}- {3\over 8 \,  C_{a,1}^2} ( C_{a,1}^2+2\,C_{a,0} \, C_{a,2})(x-e_a)\right|+ \mathcal{O}(x-e_a)^2+     \mathcal{O}(y)^2\,,
\nn\\
 b_1 = & \ \alpha'  \left[  e_1\,   - { C_{a,0} \over 4 \, C_{a,1}} + {3\over 8 \, C_{a,1}^2} ( C_{a,1}^2+2\,C_{a,0} \, C_{a,2})(x-e_a)\right] -
 b_1^0
 + \mathcal{O}(x-e_a)^2+     \mathcal{O}(y)^2\,,
\label{stringactiongen}
\end{align}
 We therefore see that branch points are minima of the action if the expansion coefficients satisfy the relation
\be
 C_{a,1}^2+2\,C_{a,0} \, C_{a,2}=0\,.
 \label{relation}
\ee
We will verify in a particular regime that this relation is satisfied. The corresponding on-shell action becomes
 \be
S_{\text{on-shell}}(e_a)=-\frac{1}{\alpha'} \left.\left( e^{\Phi\over 2} \, f^2_1+b_1\right)\right|_{v = e_a}
=- e_a+{ C_{a,0} \over 4 \, C_{a,1}}-\left|{C_{a,0}\over 4\, C_{a,1}}\right|+ \frac{b_1^0}{\alpha'}\,.
\label{stringactiong}
\ee

The general results above can be made more precise in a special limit of the underlying genus $g$ surface where the physics becomes  more transparent and a concrete expression for $w(v)$ can be proposed. In particular we consider the limit where intervals with Neumann boundary conditions or branch cuts are sufficiently far away from each other. Thus, in the surroundings of a particular branch cut, the information about the other cuts can be dismissed and $h_1$ behaves essentially as in the genus zero case. In the dual matrix model description some analogous implication will be observed for the dual resolvent function $w(v)$ in the limit where the dual Young tableau is made of large blocks.

Let us denote the $g+1$ branch cuts by $L_i$ and consider they are centered at $c_i$ and with lengths $2\mu_i$. In other words, the $2g+2$ branch points are located at $e_{2i}=c_i-\mu_i$ and $e_{2i-1}=c_i+\mu_i$. Then we propose the following expressions for $w$ over the real axis, valid for cuts well separated, {\it i.e.} $|c_i-c_j| \gg 1$. For $ x\in L_i$ or $c_i-\mu_i<x<c_i+\mu_i$
\begin{align}
w(x)= & \, 2(x-c_i)-2{\rm i}\sqrt{\mu_i^2-(x-c_i)^2} + 2 \sum_{k=1}^{i-1} \left(x-c_k+\sqrt{(x-c_k)^2-\mu_k^2}\right) \nn
\\
& + 2\sum_{k=i+1}^{g+1}\left(x-c_k-\sqrt{(x-c_k)^2-\mu_k^2}\right)\,.
\label{win}
\end{align}
While for $x$ between two cuts, {\it i.e.} $c_{i+1}+\mu_{i+1}<x<c_i-\mu_i$
\be
w(x)= 2 \sum_{k=1}^{i} \left(x-c_k+\sqrt{(x-c_k)^2-\mu_k^2}\right)
 + 2\sum_{k=i+1}^{g+1}\left(x-c_k-\sqrt{(x-c_k)^2-\mu_k^2}\right)\,.
\label{wout}
\ee
Therefore, in the vicinity of the branch cut $L_i$, and provided that $|c_i-c_j| \gg 1$, we have
\be
w(x)\approx\left\{\begin{array}{lcr}
2(x-c_i)+2\sqrt{(x-c_i)^2-\mu_i^2}\, & \quad & x<c_i-\mu_i\\
2(x-c_i)-2{\rm i}\sqrt{\mu_i^2-(x-c_i)^2}\, & \quad & c_i-\mu_i<x<c_i+\mu_i\\
2(x-c_i)-2\sqrt{(x-c_i)^2-\mu_i^2}\, & \quad & x> c_i+\mu_i \\
\end{array}\right.
\label{wnearcut}
\ee
where $\approx$ means that we are discarding terms of order $\mathcal{O}\left(\frac{1}{c_i-c_j}\right)$.
Note moreover that, when taking $x\to \infty$, we have
\be
w(x) = \frac{1}{x} \sum_{i=1}^{g+1}{\mu_i^2} +\frac{1}{x^2}\sum_{i=1}^{g+1}{c_i \mu_i^2}+ \mathcal{O}(x^{-3})\,,
\ee
thus, our requirement that $b_1$ has to vanish in the region asymptotically $AdS_5\times S^5$ implies that
\be
b_1^0 = \alpha'\frac{\sum_{i=1}^{g+1} c_i \mu_i^2 }{\sum_{i=1}^{g+1}\mu_i^2}\,.
\ee

At this point we should express the branch point parameters $\{c_i,\mu_i\}$ in terms of the brane fluxes, which are directly related to the integers
$\{n_i,k_j\}$  specifying the representation of the dual Wilson loop. These relations can be obtained from (\ref{QD3}-\ref{nkd3d5}), which gives
\begin{align}
   (4\pi^2\alpha')^2 n_i\, , =&\ 32\, \pi^2\,  {\rm i}\,  \int_{e_{2i}}^{e_{2i-1}}d\mathcal{C} + \text{c.c.} = {4 \, \pi^2 \, (\alpha')^2\over g_s} \,  {\rm i}  \int_{e_{2i}}^{e_{2i-1}}   w(x)  dx + \text{c.c.}   \, , \label{intws3}  \\
(4\pi^2\alpha') k_j =&\ -8\, \pi\, {\rm i}\,  \int_{e_{2j+1}}^{e_{2j}}d \mathcal{A}(x)+ \text{c.c.} =-{ \pi\,\alpha'\over g_s} \int_{e_{2j+1}}^{e_{2j}}d \left[ w(x)-2x \right] + \text{c.c.}  \,        \, , \label{intws5}
\end{align}
 where in the first line we  integrated by parts and used the fact that $x w(x)$ is real once evaluated at the branch points.
If we now use (\ref{wnearcut}) and since the integral is defined slightly above the real axis, we obtain
\begin{align}
n_i \approx &\ \frac{1}{2\, \pi^2 g_s} \int_{c_i-\mu_i}^{c_i+\mu_i} \sqrt{\mu_i^2-(x-c_i
)^2}=  \frac{N}{\lambda} \mu_i^2\,,
\\
k_j \approx &\ -{ 1\over 4 \pi g_s} \int_{e_{2j+1}}^{e_{2j}}d \left[ w(x)-2x \right] + \text{c.c.} =
 {4 N\over \lambda}(c_j-c_{j+1})\,.
\end{align}
We now define $\nu_i = \frac{n_i}{N}$ and $K_j = \sum_{i=j}^g k_i$, so that we can write $k_j = K_j-K_{j+1}$ and conclude that
$\mu_i = \sqrt{\lambda \nu_i}$ and $c_i = \frac{\lambda K_i}{4N}+c_0$. Since $\sum_{i=1}^{g+1}\nu_i = 1$ the gauge
fixing constant becomes
\be
b_1^0 = \alpha'\sum_{i=1}^{g+1} c_i \nu_i\,.
\ee

In order to obtain an explicit evaluation of (\ref{stringaction}) we need the coefficients $C_{a,n}$ of the expansion of
$a_0(x)$. For the proposal (\ref{wnearcut}) and for $x\in L_i$ we have
\begin{align}
a_0(x) = \frac{\alpha'}{2g_s}\sqrt{\mu_i^2-(x-c_i)^2}\,,
\label{ghexpneu}
\end{align}
Moreover, expanding around the right endpoint of the cut $x\approx e_{2i-1}=c_i+\mu_i$  we obtain an expansion of the form (\ref{a0exp}) with
\be
C_{2i-1,0}={{\rm i} \,\alpha'\over g_s} \sqrt{\mu_i \over 2} \qquad , \qquad  C_{2i-1,1}={C_{2i-1,0}\over 4\mu_i} \qquad , \qquad
C_{2i-1,2}=-{C_{2i-1,0}\over 32}\,.
\ee
 We notice that these coefficients satisfy the relation (\ref{relation}) and the on-shell string action (\ref{stringactiongen}) at the branch point reduces to
\begin{align}
S_{\rm on-shell}(e_{2i-1}) =&\ -e_{2i-1} + \frac{b_1^0}{\alpha'} = -c_i-\mu_i +\frac{1}{\lambda}\sum_{j=1}^{g+1} c_i \mu_i^2
\nn\\
=&\ -\sqrt{\lambda \nu_i} -\frac{\lambda}{4N} \left(K_i - \sum_{j=1}^{g}K_j \nu_j \right)\,.
\end{align}
Notice that going from the first to the second line, the dependence on the arbitrary constant $c_0$ cancels out, thus implying that the on-shell action is invariant under rigid translations of the branch cuts.

On the other hand, the coefficients for the expansion around the left endpoint of the cut $x\approx e_{2i}=c_i-\mu_i$  are
\be
C_{2i,0}= {\alpha'\over g_s} \sqrt{\mu_i \over 2} \qquad , \qquad  C_{2i,1} = -{C_{2i,0}\over 4\mu_i} \qquad , \qquad
C_{2i,2}= -{C_{2i,0}\over 32}\,.
\ee
They also satisfy the relation (\ref{relation}), but the on-shell string action (\ref{stringactiongen}) is in this case
\be
S_{\rm on-shell}(e_{2i})=-c_i+\mu_i -\left| \frac{C_{2i,0}}{2C_{2i,1}}\right| + \frac{b_1^0}{\alpha'} = -c_i-\mu_i + \frac{b_1^0}{\alpha'} = -e_{2i-1} + \frac{b_1^0}{\alpha'} \,.
\ee

Similar results are obtained using the expansion along the interval with Dirichlet boundary conditions. In analogy with the genus one case, configurations at the endpoints of the same brunch cut have identical on-shell actions but only $g+1$ configurations will contribute to the saddle point approximation that computes the dual correlator of Wilson loops,
\be
\boxed{
\langle W_{\rm fund} \rangle_{\mathbf{R}}\approx \sum_{i=1}^{g+1}e^{-S_{\rm on-shell}(e^*_i)} =
\sum_{i=1}^{g+1}e^{\sqrt{\lambda \nu_i} +\frac{\lambda}{4N} \left(K_i - \sum_{j=1}^{g}K_j \nu_j \right)}        ,
}
\label{correstringgenric}
\ee
where $\{e^*_i\}$ is the subset of branch points corresponding to the compatible string embeddings. For the genus one case we have seen that $\{e^*_i\}=\{e_1,e_4\}$.

As discussed above, for the correlator of Wilson loops with opposite orientations we have to change the sign in the $b_1$ contribution to the on-shell action.
Repeating the same analysis as before we obtain
\be
\langle \widetilde W_{\rm fund} \rangle_{\mathbf{R}}\approx \sum_{i=1}^{g+1}e^{-S_{\rm on-shell}(e^*_i)} =
\sum_{i=1}^{g+1}e^{\sqrt{\lambda \nu_i} -\frac{\lambda}{4N} \left(K_i - \sum_{j=1}^{g}K_j \nu_j \right)} .
\label{correstringgenric2}
\ee
\\

\section{Correlator of $\frac{1}{2}$-BPS Wilson Loops in $\mathcal{N}=4$ SYM}
\label{matrixmodel}

We now turn to the dual field theory description of the object we have been considering, {\it i.e.}, the correlator of $\frac{1}{2}$-BPS Wilson Loops in $\mathcal{N}=4$ super Yang-Mills. Specifically, we will consider the correlator of two Wilson loops
\be
\label{eq:HolCorr0}
 \langle   \,  W_\mathbf{r} \,  \rangle_\mathbf{R} =  \frac{\langle \, W_{\mathbf{R}} \,  W_\mathbf{r}  \,\rangle}{\langle \, W_\mathbf{R} \, \rangle}\,,
\ee
with the Wilson loops defined as
\be
W_{\mathbf{R}} = {\rm tr}_{\mathbf{R}} P \exp\left[\oint_{\cal C}ds\left({\rm i} A_\mu \dot x^\mu + \vec n \cdot \vec \Phi |\dot x| \right) \right]\,.
\label{wilsonloop}
\ee
The two Wilson loops in the correlator will be taken over the same circle, {\it i.e.} one on top of each other sharing the orientation in the internal space, namely be $\vec n(\tau) = \vec n_0$ with $\vec{n}_0$ a constant unitary vector in the six-dimensional internal space. By $\mathbf{R}$ and $\mathbf{r}$ we mean large and small rank representations respectively. As small representations we will successively consider the fundamental, the totally symmetric and totally anti-symmetric. We notice that the correlator  $\langle   \,  W_\mathbf{r} \,  \rangle_\mathbf{R}$ is dimensionless, and there are no other scales besides the radius of the loop, so the result should be a radius-independent function of the coupling constant.

A remarkable fact is that the expectation value of operators (\ref{wilsonloop}) is given in terms of expectation values in a Gaussian matrix model obtained through localization \cite{Pestun:2007rz}. When the rank of the representation $\mathbf{R}$ is very large, the insertion of this Wilson loop competes with the quadratic terms of the matrix model. This backreaction in the eigenvalue distribution is the field theory counterpart of the gravitational backreaction, as the dual geometry is no longer $AdS_5\times S^5$ \cite{D'Hoker:2007fq,Okuda:2008px}. This suggests
$\langle   \,  W_{\rm fund} \,  \rangle_\mathbf{R}$ should be compared with the string theory result
(\ref{correstringgenric}).

To be more specific, we are interested in computing the correlator between a Wilson loop that backreacts on the geometry and another which does not.  We are going to use the intuition of \cite{Okuda:2008px}, to first consider the correlator between backreacting Wilson loop in a representation given by a large rectangular Young tableau and a Wilson loop in the fundamental. Finally we will consider the case where the light Wilson loops is in the totally symmetric or totally antisymmetric representations by generalizing the approach of \cite{Hartnoll:2006is}. We further extend all results to the case in which the backreacting Wilson loop is in an arbitrary large representation of the gauge group.

\subsection{The back-reacting Wilson loop}

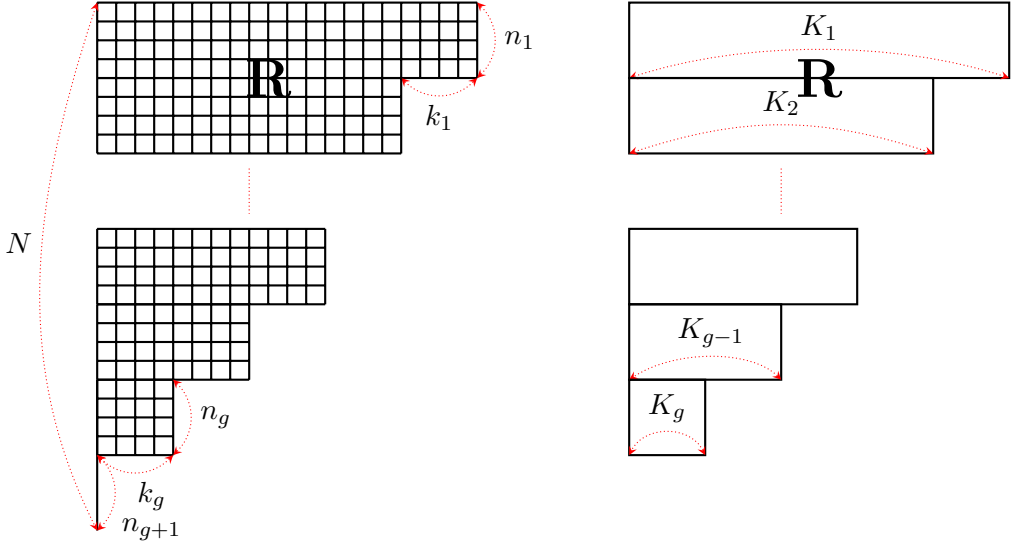
\begin{figure}[H]
\beginpgfgraphicnamed{Gen-Rep}
\begin{center}
\begin{tikzpicture}[>=stealth]
\draw[thick] (0,0) grid  [xstep=.25,ystep=.25]  (5,1);
\draw[font=\huge] (2.25,0) node {$\mathbf{R}$};
\draw[font=\huge] (9.5,0) node {$\mathbf{R}$};
\draw[thick] (7,0)rectangle (12,1);
\draw[densely dotted,<->,red] (12,0) .. controls (10.5,0.5) and (8.5,0.5) .. (7,0) node[above,pos=.5,black]{$K_1$};
\draw[densely dotted,<->,red] (5,0) .. controls (5.3,0.3) and (5.3,0.7) .. (5,1) node[right,pos=.5,black]{$n_1$};
\draw[thick] (7,-1)rectangle (11,0);
\draw[densely dotted,<->,red] (11,-1) .. controls (9.5,-0.5) and (8.5,-0.5) .. (7,-1) node[above,pos=.5,black]{$K_2$};
\draw[densely dotted,<->,red] (4,0) .. controls (4.3,-0.3) and (4.7,-0.3) .. (5,0) node[below,pos=.5,black]{$k_1$};
\draw[thick] (0,-1) grid  [xstep=.25,ystep=.25]  (4,0);
\draw[densely dotted,red] (2,-1.2)--(2,-1.8);
\draw[densely dotted,red] (9,-1.2)--(9,-1.8);
\draw[thick] (0,-3) grid  [xstep=.25,ystep=.25]  (3,-2);
\draw[thick] (7,-3)rectangle (10,-2);
\draw[thick] (0,-4) grid  [xstep=.25,ystep=.25]  (2,-3);
\draw[thick] (0,-5) grid  [xstep=.25,ystep=.25]  (1,-4);
\draw[densely dotted,<->,red] (1,-4) .. controls (1.3,-4.3) and (1.3,-4.7) .. (1,-5) node[right,pos=.5,black]{$n_g$};
\draw[thick](0,-4)--(0,-6);
\draw[densely dotted,<->,red] (1,-5) .. controls (0.7,-5.3) and (0.3,-5.3) .. (0,-5) node[below right,pos=.6,black]{$k_g$};
\draw[thick] (7,-4)rectangle (9,-3);
\draw[densely dotted,<->,red] (9,-4) .. controls (8.6,-3.6) and (7.6,-3.6) .. (7,-4) node[above,pos=.5,black]{$K_{g-1}$};
\draw[thick] (7,-5)rectangle (8,-4);
\draw[densely dotted,<->,red] (8,-5) .. controls (7.8,-4.6) and (7.2,-4.6) .. (7,-5) node[above,pos=.5,black]{$K_{g}$};
\draw[densely dotted,<->,red] (0,-5) .. controls (0.3,-5.3) and (0.3,-5.7) .. (0,-6) node[below right,pos=.7,black]{$n_{g+1}$};
\draw[densely dotted,<->,red] (0,1).. controls (-1,-2) and (-1,-4)..(0,-6) node[left,pos=0.4,black]{$N$};
\end{tikzpicture}
\end{center}
\endpgfgraphicnamed
\caption{A general representation $\mathbf{R}$ with steps given $n_i$ and $k_i$, in the right  a decomposition of the representation  in $g$  rectangles of edges $n_i$, $K_i=\sum_{j=i}^g k_j$,  all of order $N$.}
\label{fig:GenRep}
\end{figure}

In this section we review the computation of a Wilson loop in an arbitrary representation ${\bf R}$ of the gauge group \cite{Okuda:2008px}.
First, we consider the result for representations of $U(N)$ and then comment on how to obtain the result for $SU(N)$.   The expectation value of a circular Wilson loop in ${\cal N}=4$ is computed by the localization formula
\be
\langle \, W_{\bf R} \,\rangle  =\frac{1}{Z}\int  d a \, \Delta(a) \,e^{-{2N\over \lambda}  \sum_r a_r^2  }\, \tr_{\bf R} e^{  a }\,,   \label{wra}
\ee
with
\be
Z=\int  d a \, \Delta(a) \,e^{- {2N\over \lambda} \sum_r a_r^2  }\, ,
\ee
and  $da=\prod_{r=1}^N da_r$, $\Delta(a)=\prod_{r<s} (a_r-a_s)^2$ is the Vandermonde determinant and $a_r$ the eigenvalues of the matrix $a$ in the fundamental representation. A representation ${\bf R}$ of  $U(N)$ is specified by the  Dynking labels $\lambda=(\lambda_1,\lambda_2,\ldots \lambda_{N-1})$, or equivalently by a Young tableau with rows of length $\ell_r$ given by
 \be
 \ell_r=1+\sum_{s=r}^{N-1} \lambda_s   \quad\quad  r=1,\ldots N \,.
 \ee
 It is convenient to associate to any representation a Young tableau with an extra column of length $N$. We introduce the orthonormal basis $\{ e_r \}$ with
 $e_r \in \mathbb{R}^N$ and write the $U(N)$ simple roots as $\alpha_r=e_r-e_{r+1}$ for $r=1,\ldots N-1$. The character of a representation is given by the Weyl formula
  \be
 \tr_{\bf R}\, e^{  a }=\sum_{\alpha \in {\bf R} }  e^{ a\cdot \alpha }= {  {\rm det}_{r,s}     e^{ a_r (\ell_s +N-s )  }   \over   {\rm det}_{r,s}     e^{  a_r (N-s )  }   }\,,
\label{trace}
  \ee
  with the sum running over the set of weights $\{ \alpha  \}$ defining the representation ${\bf R}$.   The determinant in the numerator can be written as
  \be
   {\rm det}_{r,s}\,     e^{  a_r (\ell_s +N-s )  }  = \sum_{\sigma\in S_N}  (-1)^{\sigma}\,   \prod_{r=1}^N  e^{ a_{\sigma(r)} (\ell_{r} +N-r )  }\,,
  \ee
   while the one in the denominator can be explicitly written in the form
  \be
  {\rm det}_{r,s}   \,  e^{ a_r (N-s )  }  =\prod_{r<s}  \left(   e^{  a_r  } -   e^{  a_s  }    \right) \,.  \label{denominator}
  \ee
Alternatively the denominator can be written as
\be
\prod_{r<s}  \left(   e^{  a_r  } -   e^{  a_s  }\right) =(-1)^\sigma\prod_{r<s}  \left(   e^{  a_{\sigma(r)}  } -   e^{  a_{\sigma(s)}  }\right)\,.
\ee
with $\sigma \in S_N$ an arbitrary permutation.  Eq. \eqref{trace} can then be rewritten as
\be
\tr_{\bf R}\, e^{  a }=\sum_{\sigma\in S_N}  \,   \frac{\prod_{r=1}^N  e^{ a_{\sigma(r)} (\ell_r +N-r )  }}{\prod_{r<s}  \left(   e^{  a_{\sigma(r)}  } -   e^{  a_{\sigma(s)}  }\right)}\,.  \label{trrea}
\ee
Plugging (\ref{trrea}) into  (\ref{wra}) and renaming the dummy variables $a_{\sigma(r)} \to a_r$ one finds that any element in the sum over $\sigma$ gives the same result.
 Discarding the ${\bf R}$-independent $N!$ factor  we obtain
  \bea
\langle \, W_{\bf R} \,\rangle &=& \frac{1}{Z}\int  d a \, \Delta(a) \,e^{- {2N\over \lambda} \sum_r a_r^2  }\,
{ \prod_{r=1}^N  e^{ a_r (\ell_r +N-r )  }   \over \prod_{r<s}  \left(   e^{ a_r  } -   e^{  a_s  }    \right) } \nn\\
&=&
\frac{1}{Z}\int  d a \, \Delta(a) \,e^{ \sum_r  \left( -{ Na_r^2 \over 2\lambda}  + a_r \, \ell_r \right)    }\,
  \prod_{r<s}  \left(   1 -   e^{ a_s -a_r }      \right)^{-1}\,.
\eea
In the limit where the t'Hooft coupling $\lambda$ is large, the main contributions come from $a_r$ large, so  assuming $a_r>a_s$ for $r<s$ the exponential  terms can be dropped leading to
 \be
\langle \, W_{\bf R} \,\rangle = {1\over Z}\int  d a \, \Delta(a) \,e^{  \sum_r \left( - {2\,N\over \lambda}  a_r^2    +    a_r \, \ell_r  \right)      }\,   \label{wlr1} .
\ee
Taking the Wilson loops made of blocks of $n_i $ rows of length $K_i$ and exponentiating the Vandermonde determinant  one finds
\be
\langle \, W_{\bf R} \,\rangle = {1\over Z}\int  d a  \,\exp\left({ - {2\,N\over \lambda}  \sum_{r} a_r^2  +\sum_{r<s} \log(a_r-a_s)^2  +    \sum_{i=1}^{g+1} K_i \, \sum_{r\in {\cal I}_i} a_r
      }\right)\,   \label{wlr},
\ee
where we have split the range of $r\in \left[ 1,N\right]$ into segments ${\cal I}_i$,
 of length $n_i$, ${\cal I}_1=\left[ 1,n_1\right]$, ${\cal I}_2=\left[ n_1+1,n_1+n_2\right]$ and so on. Notice that $n_{g+1}=N-(n_1+n_2+\ldots n_g)$ and $K_{g+1}=0$.
 We display the  generic Young tableau  in Fig. \ref{fig:GenRep}.

Completing the squares in (\ref{wlr}), one can write the expectation value of the Wilson loop as
 \be
\langle \, W_{\bf R} \,\rangle = {v_{\mathbf{R}} \over Z}\int  d a   \,
\exp\left({ - {2\,N\over \lambda} \sum_i \sum_{r\in {\cal I}_i} \left( a_r - c_i \right)^2  +\sum_{r<s} \log(a_r-a_s)^2
  }\right)\, , \label{wlr2}
\ee
  with\footnote{Note that the centers $c_i$ of the matrix model branch cuts are intimately related to the centers of the branch cuts of the supergravity solution introduced in section \ref{strings} up to an arbitrary constant $c_0$ which in the matrix model is completely fixed.   }
  \be
  c_i={K_i\, \lambda\over 4\, N} \quad , \quad v_{\mathbf{R}}=\exp\left({\sum_i {n_i \, K_i^2\, \lambda\over 8 \, N}}\right)
  \ee
We are interested in the limit of large $N$  with  $K_i, n_i\approx N$. In this limit all contributions in the sum are of  order $N^2 $ and cannot be dropped when using the saddle point approximation.  The saddle point equations then read
\begin{eqnarray}
\label{eq:saddleq}
&&-\frac{4 N}{\lambda} (a_r-c_i)+2\sum_{s\neq r} \frac{1}{ a_r-a_s}  =0\, ,   \qquad   r\in {\cal I}_i\,,
\end{eqnarray}
  or in its continuous version\footnote{Here $\rho(x)={1\over N} \sum_r \delta(x-a_r)$. }
  \be
  -\frac{4 N}{\lambda}  (x-c_i) +2 N  \int  dy \, {\rho(y) \over x-y }   =0\,,    \qquad        c_i-\mu_i < x < c_i+\mu_i\,,   \label{eqsaddle}
  \ee
  with $\mu_i>0$ some real numbers. These equations are solved \cite{Okuda:2008px} by taking the matrix model resolvent $w(x)$
    \be
  w(z)=\lambda \int_{-\infty}^\infty {\rho(y) \over z-y}\,,
  \ee
  to be given by the integral
  \be
  w(z)=\int_{\infty}^z \alpha\,,
  \label{doble v}
  \ee
   of a meromorphic one form
  \be
   \alpha(z) =2\left(  1-{a_{g+1}(z)\over \sqrt{H_{2g+2}(z)} } \right)\,,
   \label{alfa}
   \ee
  defined  on the hyperelliptic curve $y^2=H_{2g+2}(z) $ with $H_{2g+2}(z)$ and $a_{g+1}(z)$ polynomials of order $2g+2$ and $g+1$ respectively.  The parameters specifying these polynomials are uniquely given in terms of $K_i$ and $n_i$. By considering integrals of \eqref{doble v} and \eqref{alfa} over non-trivial cycles on the hyperelliptic surface, one finds constraints analogous to the expressions \eqref{intws3} and \eqref{intws5} giving the supergravity charges of the dual bubbling geometry. Then, it is natural   to identify the matrix model resolvent with the holomorphic function introduced in \eqref{AB} as proposed   in  \cite{Okuda:2008px}.
  %

   \subsubsection{ Multi-cut Wigner semicircle distribution}

   To make an explicit comparison with string theory results, here we focus on the case where  the distances between the  cuts are large. First, we observe that for a single cut, (\ref{eqsaddle}) is solved by taking $ \rho(y)={2\over \pi \, \mu }\, \sqrt{ \mu^2-y^2}$.
    In the limit where the interactions between the  eigenvalues within different intervals can be neglected, the solution to (\ref{eqsaddle}) can be found as\footnote{ Note this eigenvalue distribution is in complete agreement with the proposed gravity solution in terms of the $w$ function \eqref{win},\eqref{wout}, if we further identify this function with the resolvent of the matrix model, namely
\be
w(z)=\lambda\int \frac{\rho(y)}{z-y}\approx \frac{2}{\pi}\sum_{i=1}^{g+1}\int_{c_i-\mu_i}^{c_i+\mu_i}\frac{\sqrt{\mu_i^2-(y-c_i)^2}}{z-y}
\ee  }
  \be
\rho(x)= \Bigg \{ \begin{array}{ccc}
    {2\over \pi \, \lambda }\, \sqrt{ \mu_i^2-(x-c_i)^2}    & , &  c_i - \mu_i < x < c_i + \mu_i, \\
0 & , & {\rm otherwise}\, ,
\label{twocutdensity}
\end{array}
\ee
with  centers and half-lengths given by
  \bea
c_i &=&{K_i\, \lambda\over 4\, N}  \qquad , \qquad   \mu_i=\sqrt{\lambda \nu_i }   \qquad  {\rm for} \qquad i=1,\ldots g+1 \nn\\
K_{g+1} &=& 0      \qquad , \qquad   \nu_{g+1}= 1-\sum_{i=1}^g \nu_i \,,
  \eea
where we have defined $\nu_i=\frac{n_i}{N}$ and normalised the eigenvalues distributions as
\be
\int_{c_i-\mu_i}^{c_i+\mu_i} \rho(x) \,dx=\nu_i
\ee
Finally, the expectation value \eqref{wlr2} evaluated in the multi-cut  eigenvalue distribution reduces to
\be
\left\langle W^{U(N)}_{\mathbf{R}} \right\rangle \approx  \exp\left(\frac{\lambda}{8N} \sum_{i=1}^g n_i\, K_i^2\right)\,,
\label{matrix2}
\ee
where $\approx$ here implies we are discarding subleading contributions of order $N^2\log\lambda$.

In the case of $SU(N)$  there is an additional factor of $(\det(e^M))^{-\frac{|\mathbf{R}|}{N}}$ in the matrix model integral with
$ |\mathbf{R}|= N \sum_{i=1}^g K_i\, \nu_i $. This insertion results simply into
 a rigid shift of  all centers by  $-\frac{|\mathbf{R}|\, \lambda}{4N^2} $  or equivalently
\be
K_i \to  K_i- \sum_{j=1}^g K_j\, \nu_j\,.
\label{eq:shift}
\ee
For the expectation value of the Wilson loop one finds
\be
\left\langle\,  W^{SU(N)}_{\mathbf{R}} \, \right\rangle \approx    \exp\left({\frac{\lambda}{8N} \sum_{i=1}^g n_i\, \left( K_i-\sum_{j=1}^g K_j\, \nu_j     \right)^2}\right)\,.
\label{matrix2SU}
\ee
After having reviewed the distribution of eigenvalues found in \cite{Okuda:2008px}, we proceed to compute correlators with other Wilson loops, by evaluating expectation values of appropriate insertions. We will first consider the correlator with a fundamental Wilson loop and then move to the cases of correlators with  totally symmetric and anti-symmetric Wilson loops.

 \subsection{Adding a fundamental Wilson loop}

 Computing the correlator between a large Wilson loop and a Wilson loop in the  fundamental representation translates in the matrix model to evaluating the expectation value of the operator $\sum_{r=1}^N e^{a_r}$ in the matrix model integral  \eqref{wlr2}
\bea
\langle \, W_{\bf R} \,W_{\rm fund} \,\rangle  &=& \frac{1}{Z}\int  d a \, \Delta(a) \,e^{-{2N\over \lambda}  \sum_r a_r^2  }\, \tr_{\bf R}\, e^{  a }\, \tr_{\rm fund} \,e^{  a }\,, \nn\\
&=& {v_{\bf R} \over Z}\int  d a  \sum_{i=1}^{g+1} \sum_{r\in {\cal I}_i} \,
e^{-S_r}
\eea
with
\be
S_r=  {2\,N\over \lambda} \sum_{i=1}^{g+1} \sum_{s\in {\cal I}_i} \left( a_s - c_i \right)^2  -\sum_{s<t} \log(a_{s}-a_{t})^2   -a_r\,,
\ee
This insertion is not back-reacting  in the sense that it does not modify the $\rho$-distribution discussed in the previous subsection. Taking the ratio with
 $\langle \, W_{\bf R}  \,\rangle $, the factor $v_{\bf R}$ cancels between numerator and denominator, and after the large $N$ limit one finds
 \be
 \langle \,  \,W_{\rm fund} \,\rangle_{\bf R} = \int_{-\infty}^\infty \, dx\, \rho(x)\, e^x \approx {2\over \pi \, \lambda }\sum_{i=1}^{g+1}   \int_{c_i-\mu_i}^{c_i+\mu_i} \, dx\,  \, \sqrt{ \mu_i^2-(x-c_i)^2}\, e^x\,,
  \label{intrho}
 \ee
   where $\approx$ denotes the approximation where  centers are far away from each other, {\it i.e.} $K_i -K_j \gg  N$ and the interactions between the regions ${\cal I}_i$ have been neglected.  By doing the integrals we get the typical Bessel functions,
\be
\label{eq:WrectaWf}
\langle \,  \,W_{\rm fund} \,\rangle_{\bf R} \approx \sum_{i=1}^{g+1} \frac{2 \mu_i}{ \lambda}\, e^{c_i}\, I_1(\mu_i)\approx\sum_{i=1}^{g+1} e^{c_i+\mu_i}\,.
\ee

For comparison with the string theory results in the context of the AdS/CFT correspondence, we should focus on the $SU(N)$ matrix model. In that case

\be
\boxed{\langle \,  \,W_{\rm fund}^{SU(N)} \,\rangle_{\bf R}
\approx \sum_{i=1}^{g+1} e^{\sqrt{\lambda\nu_i}+\frac{\lambda}{4N}\left(K_i-\sum_{j}^{g}K_j \nu_j\right)}}\,.
\label{eq:WrectaWfsun}
\ee
that matches precisely the  AdS/CFT prediction \eqref{correstringgenric}.

For instance, in the case of a representation given by a rectangular Young tableau, the position of the centers are
\bea
 c^{SU(N)}_1 &=&  { k\,\lambda\over 4\, N} (1-\nu)\,, \qquad  \qquad c^{SU(N)}_2=- { k\,\nu\,\lambda\over 4\, N}\,,
 \eea
and (\ref{eq:WrectaWfsun}) yields
\be
\langle \,  \,W_{\rm fund}^{SU(N)} \,\rangle_{\bf R} \approx    e^{\sqrt{\nu\,\lambda}+\frac{k(1-\nu) \lambda}{4N}}   +    e^{\sqrt{\lambda(1-\nu)}- { k\,\nu\,\lambda\over 4\, N} }     \, ,
\ee
that matches the result \eqref{eq:HolCorr}.

Before moving to correlators in more general representations, let us consider the correlator with another fundamental Wilson loop that can also be computed with the matrix model. At the end of section \ref{strings} we considered the possibility of a correlator of two loops with opposite spatial orientations. It turns out, as shown in appendix \ref{susywl}, that if the internal orientation is also opposite, the two loops are invariant under the same set of supersymmetries and therefore their correlator can be accounted for by an expectation value in the Gaussian matrix model. Since the internal space orientation is opposite, the matrix model computation is in this case
\be
 \langle \,  \,\widetilde{ W}_{\rm fund} \,\rangle_{\bf R} \approx \int_{-\infty}^\infty \, dx\, \rho(x)\, e^{-x}\,.
  \label{intrho2}
 \ee
For the case of the $SU(N)$ matrix model, we get now
\begin{align}
\langle \,  \,\widetilde{W}_{\rm fund}^{SU(N)} \,\rangle_{\bf R}
&\approx  \sum_{i=1}^{g+1} \frac{2 \mu_i}{ \lambda}\, e^{-c_i}\, I_1(\mu_i)\approx\sum_{i=1}^{g+1} e^{-c_i+\mu_i} \nn\\
&\approx \sum_{i=1}^{g+1} e^{\sqrt{\lambda\nu_i}-\frac{\lambda}{4N}\left(K_i-\sum_{j}^{g}K_j \nu_j\right)}\,.
\label{eq:WrectaWfsun2}
\end{align}
Once again this is in agreement with the AdS/CFT prediction \eqref{correstringgenric2}.
If we restrict ourselves to the case of a representation given by a rectangular Young tableau, the result becomes
\be
\langle \,  \,\widetilde{W}_{\rm fund}^{SU(N)} \,\rangle_{\bf R} \approx    e^{\sqrt{\nu\,\lambda}-\frac{k(1-\nu) \lambda}{4N}}   +    e^{\sqrt{\lambda(1-\nu)}+ { k\,\nu\,\lambda\over 4\, N} }\,,
\ee
thus matching the explicit result \eqref{eq:HolCorr2}.

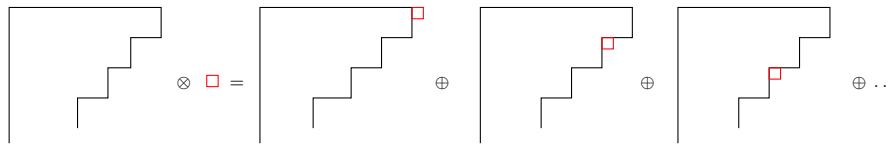
\begin{figure}[H]
\beginpgfgraphicnamed{Tensor-Prod}
\begin{center}
\begin{tikzpicture}[>=stealth]
\path[draw] (4.2,-0.8)-- (4.2,1)--(6.2,1)--(6.2,0.6)--(5.8,0.6)--(5.8,0.2)--(5.5,0.2)--(5.5,-0.2)--(5.1,-0.2)--(5.1,-0.6);
\draw[font=\tiny] (6.5,0) node {$\otimes$};
\draw[red]  (6.8,-0.05) rectangle (6.95,0.1);
\draw[font=\tiny] (7.2,-0.02)node{$=$}(7.2,-0.02);
\path[draw] (7.5,-0.8)-- (7.5,1)--(9.5,1)--(9.5,0.6)--(9.1,0.6)--(9.1,0.2)--(8.7,0.2)--(8.7,-0.2)--(8.2,-0.2)--(8.2,-0.6);
\path[draw,red] (9.5,1)--(9.65,1)--(9.65,.85)--(9.5,.85)--(9.5,1);
\draw[font=\tiny] (9.9,0.0) node {$\oplus$};
\path[draw] (10.4,-0.8)-- (10.4,1)--(12.4,1)--(12.4,0.6)--(12.0,0.6)--(12.0,0.2)--(11.6,0.2)--(11.6,-0.2)--(11.2,-0.2)--(11.2,-0.6);
\path[draw,red] (12.0,0.6)--(12.15,0.6)--(12.15,0.45)--(12.0,.45)--(12.0,0.6);
\draw[font=\tiny] (12.6,0.0) node {$\oplus$};
\path[draw] (13.0,-0.8)-- (13.0,1)--(15.0,1)--(15.0,0.6)--(14.6,0.6)--(14.6,0.2)--(14.2,0.2)--(14.2,-0.2)--(13.8,-0.2)--(13.8,-0.6);
\path[draw,red] (14.2,0.2)--(14.35,0.2)--(14.35,0.05)--(14.2,.05)--(14.2,0.2);
\draw[font=\tiny] (15.6,0.0) node {$\oplus$\,\,\dots};
\end{tikzpicture}
\end{center}
\endpgfgraphicnamed
\caption{Tensor product between a `large' representation and a fundamental one. }
\label{fig:ProdYT}
\end{figure}

So far we have computed correlators of Wilson loops defined over coincident circular contours. This amounted to compute the expectation value of the product ${\rm tr}_{\bf R}e^M {\rm tr}_{\bf r}e^M$.  However, there is an alternative and interesting point of view, which arises from the ring structure of the characters of the gauge group representations, namely 
\bea
\tr_{\mathbf{R}} e^{M}\tr_{\mathbf{r}} e^{M}=\tr_{\mathbf{R}\otimes\mathbf{r} } e^{M}=\sum_{\mathbf{R}_i\in \text{irreps}}C_{\mathbf{R}\mathbf{r}\mathbf{R}_i} \, \tr_{\mathbf{R}_i}e^{M},
\eea
where $C_{\mathbf{R}\mathbf{r}\mathbf{R}_i}$ are the multiplicities and ``irreps'' denote the irreducible components of ${\bf R}\otimes {\bf r}$. For the products we have considered in this section, $\mathbf{R}$  is  a `large'  back-reacting representation associated to a Young diagram made of $g$ blocks and  $\mathbf{r}$ is the fundamental one. In this case, the decomposition is rather simple, leading to a sum of $g+1$ irreps all of them with multiplicities $C_{\mathbf{R}\mathbf{r}\mathbf{R}_i}$ equal to 1, as schematically depicted in Figure \ref{fig:ProdYT}.

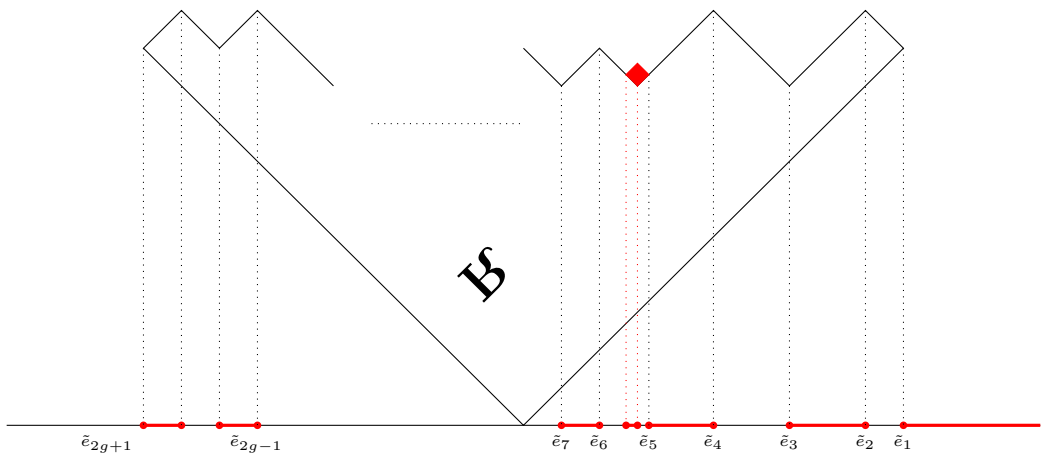
\begin{figure}[H]
\beginpgfgraphicnamed{Gen-Rep}
\begin{center}
\begin{tikzpicture}[>=stealth]
\node[yscale=-1,inner sep=0,outer sep=0,rotate=-45]  at (-0.5,2.0) {\huge $\mathbf{R}$};
\path[draw](-6.8,0)--(6.8,0);
\coordinate [label=below left: {\tiny $\tilde{e}_{2g+1}$}] (2g+1) at (-5,0);
\coordinate [] (2g) at (-4.5,0);
\coordinate [label=below right: {\tiny $\tilde{e}_{2g-1}$}] (2g-1) at (-4,0);
\coordinate [] (2g-2) at (-3.5,0);
\coordinate [label=below: {\tiny $\tilde{e}_{7}$}] (7) at (0.5,0);
\coordinate [label=below: {\tiny $\tilde{e}_{6}$}] (6) at (1,0);
\coordinate [label=below: {\tiny $\tilde{e}_{5}$}] (5) at (1.65,0);
\coordinate [label=below: {\tiny $\tilde{e}_{4}$}] (4) at (2.5,0);
\coordinate [label=below: {\tiny $\tilde{e}_{3}$}] (3) at (3.5,0);
\coordinate [label=below: {\tiny $\tilde{e}_{2}$}] (2) at (4.5,0);
\coordinate [label=below: {\tiny $\tilde{e}_{1}$}] (1) at (5,0);
\draw[red,very thick] (2g+1)--(2g);
\draw[red,very thick] (2g-1)--(2g-2);
\draw[red,very thick] (7)--(6);
\draw[red,very thick] (5)--(4);
\draw[red,very thick] (3)--(2);
\draw[red,very thick] (1)--(6.8,0);
\draw[red,very thick] (1.35,0)--(1.5,0);
\fill[red] (-5,0) circle (0.05cm);
\fill[red] (-4.5,0) circle (0.05cm);
\fill[red] (-4,0) circle (0.05cm);
\fill[red] (-3.5,0) circle (0.05cm);
\fill[red] (1,0) circle (0.05cm);
\fill[red] (1.65,0) circle (0.05cm);
\fill[red] (2.5,0) circle (0.05cm);
\fill[red] (3.5,0) circle (0.05cm);
\fill[red] (4.5,0) circle (0.05cm);
\fill[red] (5,0) circle (0.05cm);
\fill[red] (1.35,0) circle (0.05cm);
\fill[red] (0.5,0) circle (0.05cm);
\fill[red] (1.5,0) circle (0.05cm);
\path[draw](-2.5,4.5)--(-3.5,5.5)--(-4.0,5.0)--(-4.5,5.5)--(-5,5)--(0,0)--(5,5)--(4.5,5.5)--(3.5,4.5)--(2.5,5.5)--(1.5,4.5)--(1.0,5)--(0.5,4.5)--(0.0,5);
\draw[dotted](-5,5)--(-5,0);
\draw[dotted](-4.5,5.5)--(-4.5,0);
\draw[dotted](-4.0,5.0)--(-4,0);
\draw[dotted](-3.5,5.5)--(-3.5,0);
\draw[dotted](-2,4.0)--(0,4.0);
\draw[dotted](5.0,5.0)--(5,0);
\draw[dotted](4.5,5.5)--(4.5,0);
\draw[dotted](3.5,4.5)--(3.5,0);
\draw[dotted](2.5,5.5)--(2.5,0);
\draw[dotted](1.65,4.65)--(1.65,0);
\draw[dotted,red](1.5,4.5)--(1.5,0);
\draw[dotted](1.0,5)--(1,0);
\draw[dotted](0.5,4.5)--(0.5,0);
\path[draw,red,fill] (1.5,4.5)--(1.65,4.65)--(1.5,4.8)--(1.35,4.65)--(1.5,4.5) ;
\draw[dotted,red](1.35,4.65)--(1.35,0);
\end{tikzpicture}
\end{center}
\endpgfgraphicnamed
\caption{One of the diagrams depicted in figure \ref{fig:ProdYT} of a general bubbling geometry with an additional box in red. From the gravity side, the additional red box  corresponds to the collapse of one branch cut in a genus $g+1$ geometry. This is pictorially interpreted as the additional red cut that collapses  and  approaches $\tilde{e}_5$ in the figure.}
\label{fig:GenYTplusbox}
\end{figure}

Note that this exactly coincides with the number of saddles points we considered in our string theory  computation, and also with the number of contributions that appeared in the matrix model computation. This field theory remark also suggests and interpretation  for each saddle point contribution in string theory, as coming from a $g+1$ bubbling solution where one of the  branch cuts is collapsing (see figure \ref{fig:GenYTplusbox}).

\subsection{Small loops in symmetric or antisymmetric representations}

In this  section we consider other examples of correlators of a backreacting rectangular Young tableau representation Wilson loop with  non backreacting Wilson loops in  the totally symmetric and totally antisymmetric representation. We  write
\be
 \langle W_{\mathbf{r}} \, \rangle_{\mathbf{R}} = \frac{\langle\, W_{\mathbf{R}}\, W_{\mathbf{r}}  \, \rangle}{\langle \,W_{\mathbf{R}}\, \rangle}
=\int_{-\infty}^\infty \rho(x)  \Omega_{\mathbf{r}}(x)\,,
\label{correlatorO}
\ee
where $\Omega_{\mathbf{r}}(x)$ is some function corresponding to the insertion $W_{\mathbf{r}}$ in the continuous large $N$ limit and the eigenvalue distribution, $\rho(x)$, is given by the two-cuts case of (\ref{twocutdensity}). We stress, again, that this distribution is reliable in the limit where both semicircles are sufficiently far away from each other, that is, when $\frac{k\lambda}{4N}$ is sufficiently large.

The normalized correlator with (anti)-symmetric Wilson loops can be written compactly using the generating function of characteristic polynomials as in \cite{Hartnoll:2006is}:
\be
\langle W_{S_l,A_l}\rangle_{\mathbf{R}}
=
\frac{1}{{\rm dim}_{S_l,A_l}}
\oint_\Gamma \frac{dt}{2\pi\, {\rm i}} \frac{1}{t^{l+1}} \exp\Big(\mp N \int_{-\infty}^\infty dx \rho(x)\log(1\mp\, t\, e^x)\Big)\,,
\label{generating}
\ee
where we take the $-$ sign for the totally symmetric representation, $S_l$, and the $+$ sign for the  totally anti-symmetric representation, $A_l$.
The  contour $\Gamma$ encloses the pole at $t=0$. We want to evaluate the integral (\ref{generating}) for large $N$, and for a general (anti)-symmetric representation even when $l$ is large, but  not as large that can possibly back-react on the eigenvalue distribution.

\subsubsection{Correlator with a totally symmetric Wilson loop}
\label{sec:Symm}
We start by considering the totally symmetric case. We have to evaluate the integral \eqref{generating} for the  two-cut density distribution (\ref{twocutdensity}).
 It is convenient to change variables  $x \to  c_i - a_i x$ along each cut ${\cal I}_i$ in such a way as to bring the $x$-integrals to the intervals $\left[ -1,1\right]$
 \be
 \int_{-\infty}^\infty dx \rho(x)\log(1-\, t\, e^x) = \sum_{i=1}^2 \mu_i \, \int_{-1}^1 \sqrt{1-x^2}  \log(1-e^{-\mu_i x+c_i} t)\,.
 \ee
 It is also convenient to change the $t$ variable,  $t=e^{z}$, which yields
 \begin{align}
 \oint_{\tilde{\Gamma}}dz
 \exp\!& \left[  -N \left( \sum_{i=1}^2 \tfrac{2\, \mu_i^2}{\pi\lambda}\!\int\limits_{-1}^{1}\!dx \sqrt{1-x^2}\log(1-e^{-\mu_1 x+c_i+ z} )   +f\, z\right) \right]\,,
 \end{align}
 where $f=\frac{l}{N}$. The integral above  has two branch cuts  in  $z$ due to the $\log$. They are given by
 \bea
 \label{eq:zcuts1}
{-\mu_i-c_i}\leq z\leq {\mu_i-c_i}\quad \text{with }\quad   i=1,2\,.
 \eea
 The contour $\tilde{\Gamma}$ is picking  now the pole at infinity, so it  can be deformed  to pass  just above  and below the cuts. Using Jordan Lemma the  contour integral reduces to the discontinuity across the cuts of the integral:
{\small
 \begin{align}
 \label{eq:ImWL}
 & \langle W_{S_l} \rangle_{\mathbf{R}} \approx \frac{1}{\pi} {\rm Im}
 \Bigg\{   \sum_{j=1}^2
\int \limits_{-c_j-\mu_j}^{-c_j+\mu_j} \! \!dz
\exp\Bigg[  -{N \over \lambda}  \Big( \sum_{i=1}^2
 \frac{2 \, \mu^2_i}{\pi } \int\limits_{-1}^{1}\!dx\sqrt{1-x^2}
 \log(1 -e^{-\mu_i x+c_i   +  z  })  +f\, \lambda\, z\Big) \Bigg]  \Bigg\}\\
 &=   {\rm Im}
 \Bigg\{   \sum_{j=1}^2 \, \frac{\mu_j}{\pi} \,
\int \limits_{-1 }^{1 } \! \!dz
\exp\Bigg[  -{N \over \lambda} \Big( \sum_{i=1}^2
 \frac{2 \, \mu^2_i}{\pi} \int\limits_{-1}^{1}\!dx\sqrt{1-x^2}
 \log(1 -e^{-\mu_i x+c_i-c_j+ \mu_j z})  +f \, \lambda\, (\mu_j\,z-c_j  ) \Big) \Bigg]  \Bigg\}\,,\nn
\end{align}}
where in the second line we made the change of variables  $ z\to \mu_j z-c_j $. The $x$-integrals here are formal because the integrand has branch cuts along the  integration region. A way to cure this is to give  $z$ a small imaginary part ${\rm i}\, \epsilon$, so we are passing through a line slightly above the real axis.   The integrals  for $j=1,2$ can be  evaluated separately using the large-$N$ saddle point method.
The $z$-integral is dominated by the region $z\approx z^*$  extremizing the exponential term.  Let us consider the case $j=1$ and take $c_1-c_2\gg 1$. In this limit, only the $i=1$ term in the sum contributes.
To compute the saddle equations
it is convenient to break the $x$-integral   into pieces such that the argument of the log is always positive. We write
\bea
\int\limits_{-1}^{1}\!dx\sqrt{1-x^2}
 \log(1 -e^{\mu_1 (z-x) }) &=& \int\limits_{-1}^{z}\!dx\sqrt{1-x^2}
 \log(e^{\mu_1 \, z} -e^{\mu_1 x }) + \int\limits_{z}^{1}\!dx\sqrt{1-x^2}
 \log(e^{\mu_1 \, x} -e^{\mu_1 z }) \nn\\
 && + {\rm i} \, \pi \, \int\limits_{-1}^{z}\!dx\sqrt{1-x^2}\,.  \label{decomp}
\eea
We are going to look for solutions when $\text{Re}(z)<-1$ in this case the saddle equation becomes\footnote{For  $\text{Re}(z)>1$ there are no solutions to the saddle equation.}
\be
 { 2\, \mu_1\,  \over \pi}  \int\limits_{-1}^{1}\!dx
  \frac{\sqrt{1-x^2}}{ 1- e^{\mu_1 (x-z) }}
  +4\, {\rm i}\,  \mu_1 \sqrt{1-z^2} + \lambda \, f =0. \label{sad0}
 \ee
In this domain,
 the  integral term in Eq. (\ref{sad0}) can be discarded when  $\mu_1$ is large and the saddle point equation  reduces to
\be
  4\, {\rm i}\,  \mu_1 \sqrt{1-z^2} + \lambda \, f =0 \label{sad00}\,,
 \ee
with solution
 \be
 z^{*}=-\sqrt{1+\kappa_1^2}\,, \qquad\text{with}\qquad\kappa_1=\frac{f\lambda}{2 \mu_1}= {l \over 2N} \sqrt{\lambda\,    \over   \nu }\,.
 \ee
   Evaluating (\ref{decomp}) at the saddle point $z^*$  and discarding $e^{\mu_1 z}$-terms inside of the logs  one finds\footnote{The integral is computed using
 \be
{\rm i} \int_{-1}^z \sqrt{1-x^2} dx=-\int_0^{{\rm arccosh}\, z}   \sinh^2 y \, dy=\frac{1}{2} (y-\sinh y \, \cosh y )\Big|_0^{{\rm arccosh \, z } } =
\frac{1}{2}  \left({\rm arccosh \, z }-z\sqrt{z^2-1} \right)
 \ee
 }
 \bea
 \label{eq:I11}
\int\limits_{-1}^{1}\!dx\sqrt{1-x^2}
 \log(1 -e^{\mu_j (z^*-x) }) &  \underset{ \mu_i \to \infty}{\approx}   &   2{\rm i} \, \pi \, \int\limits_{-1}^{z^*}\!dx\sqrt{1-x^2} \nn\\
 &=& \pi  \left({\rm arccosh \, z^* }-z^*\sqrt{(z^*)^2-1} \right)
 \eea
 To get the contribution from this saddle point we need to evaluate the exponential in \eqref{eq:ImWL} at $z^*$. Strictly speaking this  quantity is not well defined due to the branch cuts of the exponent and for that we have added an small imaginary part to $z$, so, we will do the same for $z^*$, indeed, the well defined quantity is the imaginary part \eqref{eq:ImWL}, we essentially  need to evaluate the right hand side of \eqref{eq:I11}  taking into account this imaginary shift, and evaluate the full answer with this small  deformation.
 Taking $z^*=-\sqrt{1+\kappa_1^2}+{\rm i} \epsilon$ one finds
 \be
 \label{eq:I1}
{\rm i} \,\int\limits_{-1}^{z^*}\!dx\sqrt{1-x^2} =  \frac{1}{2} \left(\kappa_1  \sqrt{1+\kappa_1 ^2}-{\rm arcsinh}\, \kappa_1 \right)+\text{imaginary part}\,.
\ee

Plugging the solution into  (\ref{eq:ImWL}) one finally finds for the contribution of  the first saddle point
  \be
  \langle W_{S_l} \rangle_{\mathbf{R}}^{(1)} \approx
  \exp\Bigg[
 \frac{2\, N\, \mu_1^2}{\lambda } \left(\kappa_1\sqrt{1+\kappa_1^2} + \text{arcsinh}\, \kappa_1\right)+ N\, f\, c_1
  \Bigg]\,,
\ee
  where we have discarded a large $N$ phase in the result above.  For $j=2$, one follows the same steps but now  we have an extra contribution coming from the term with $i \neq j$ given by
\be
 \frac{2 \, \mu^2_1}{\pi} \int\limits_{-1}^{1}\!dx\sqrt{1-x^2}
 \log(1 -e^{-\mu_1 x+c_1-c_2+ \mu_2 z})   \approx      \mu^2_1\,    ( c_1-c_2+ \mu_2 z)\,.
\ee
The saddle point equation now becomes
\be
4\, {\rm i}\,  \mu_2 \sqrt{1-z^2} +   \lambda \, f  +    \mu^2_1  =0\,.
\ee
The solution is now given by
\bea
z^*=-\sqrt{1+\kappa_2^2}, \quad \text{with} \quad\kappa_2=\frac{\lambda f}{4 \mu_2}+\frac{\mu_1^2}{4\, \mu_2}=\frac{f\sqrt{\lambda}}{4\sqrt{1-\nu}}+\frac{\sqrt{\lambda}}{4}\frac{\nu}{\sqrt{1-\nu}}\,.
\eea
Plugging this into (\ref{eq:ImWL})
\bea
\langle W_{S_l}\rangle_{\mathbf{R}}^{(2)} \approx \exp\Bigg[\frac{2N \mu_2^2}{\lambda } \left(\kappa_2\sqrt{1+\kappa_2 ^2} + \text{arcsinh}\, \kappa_2\right)+ N(1+f)c_2  \Bigg]\,,
\eea
where the $N c_2$ term comes from the extra term $-N\, \mu_1^2(c_1-c_2)/\lambda$.  Finally, the total contribution to the correlator with the $S_l$ representation adds up to,
\begin{align}
\langle W_{S_l}\rangle_{\mathbf{R}} \approx &\exp\Bigg[2N(1- \nu) \left(\kappa_2\sqrt{1+\kappa_2 ^2} + \text{arcsinh}\, \kappa_2-\frac{1+f}{1-\nu}\frac{k\lambda}{8N}\nu\right) \Bigg]\nonumber\\&
+\exp\Bigg[2N \nu \left(\kappa_1\sqrt{1+\kappa_1 ^2} + \text{arcsinh}\, \kappa_1+f\frac{1-\nu}{\nu}\frac{k\lambda}{8N}\right)\Bigg].
\label{symfinal}
\end{align}

A comment is in order, in \cite{Hartnoll:2006is}  there was an additional solution to the saddle point equations which in the large $\lambda$ regime and $\kappa_i$ fixed or $\frac{l}{N}$ fixed, was sub-leading with respect to the contribution of the saddle point considered here. We report these contributions in the Appendix \ref{sec:2ndsad}.

\subsubsection{Correlator with a totally anti-symmetric Wilson loop}

Let us now turn our attention to the correlator with a Wilson loop in a totally anti-symmetric representation which is given by \eqref{generating}
with the two-cut distribution given in \eqref{twocutdensity}.
Performing the transformation $t=e^{\mu_2z-c_2}$ and defining $f=\frac{l}{N}$ the integral above can be rewritten as
\begin{align}
\langle W_{A_l}\rangle_{\mathbf{R}}\approx\int_{\tilde{\Gamma}} dz\exp\Big[& \frac{2N}{\lambda \pi}\Big( \mu_2^2 \int\limits_{-1}^{1} dx \sqrt{1-x^2}\log\left(1+ e^{-\mu_2(x-z)}\right)\nn\\
+&\mu_1^2 \int\limits_{-1}^{1} dx \sqrt{1-x^2}\log\left(1+ e^{-\mu_1x+\mu_2z+(c_1-c_2)}\right)-\frac{\pi\lambda }{2}(\mu_2z-c_2)f\Big)\Big].
\label{antisym}
\end{align}
Note that the branch cuts of the integrand are now along the horizontal segments $[-1+{\rm i}\,\pi,1+{\rm i}\,\pi]$ and $[-\frac{1}{\mu_2}(c_1-c_2)-\frac{\mu_1}{\mu_2}+{\rm i}\,\pi,-\frac{1}{\mu_2}(c_1-c_2)+\frac{\mu_1}{\mu_2}+{\rm i}\,\pi]$, together with the images obtained by shifting the imaginary part by multiples of $2\pi$. As in the symmetric case, we deform the contour $\tilde{\Gamma}$ to lay along the real axis and approximate the integral by its large $N$ saddle point. Unlike the previous case, the saddle point value is not located over any branch cut, making the evaluation much more straightforward. The saddle point equation reads
\be
\mu_2^2\int_{-1}^{1} dx \frac{\sqrt{1-x^2}}{1+ e^{\mu_2(x-z)}}+\mu_1^2\int_{-1}^{1} dx \frac{\sqrt{1-x^2}}{1+ e^{\mu_1x-\mu_2z-(c_1-c_2)}}-\frac{\pi\lambda }{2}f=0\,.
\label{antisymeom}
\ee
Now we search for solutions in the large $\mu_i$ regime. It turns out that the solutions can only be placed along the segments $[-1,1]$ and $[-\frac{1}{\mu_2}(c_1-c_2)-\frac{\mu_1}{\mu_2},-\frac{1}{\mu_2}(c_1-c_2)+\frac{\mu_1}{\mu_2}]$. Otherwise, the integrals in \eqref{antisymeom} become $z$-independent thus not having any solution there.

Let us first consider the region $-1<z<1$. Taking into account that $c_1-c_2=\frac{k\lambda}{4N}\gg 1$, equation \eqref{antisymeom} reduces to
\be
\mu_2^2\int_{-1}^z dx\sqrt{1-x^2}+\frac{\pi \mu_1^2}{2}-\frac{\pi\lambda }{2}f=0\,,
\ee
which yields
\be
\text{arccos}(z)-z\sqrt{1-z^2}=\pi\left(1+\frac{\mu_1^2}{\mu_2^2}-\frac{\lambda}{\mu_2^2}f\right)\,.
\ee
The solution is $z=\cos \theta_{2}$ with $\theta_{2}$ such that
\begin{align}
\theta_{2}-\sin \theta_{2}\cos\theta_{2}&=\pi\left(1+\frac{\mu_1^2}{\mu_2^2}-\frac{\lambda}{\mu_2^2}f\right) = \pi \left(1+\frac{\nu}{1-\nu}-\frac{l}{N(1-\nu)}\right)\,,
\label{theta2}
\end{align}
and then  the integral \eqref{antisym} results in
\begin{align}
\langle W_{A_l}\rangle_{\mathbf{R}}^{(2)}&\approx\exp\Big[\frac{2N}{\lambda\pi}\left(\mu_2^3\int_{-1}^{\cos\theta_2} dx\, x\sqrt{1-x^2}+\frac{\pi \mu_1^2}{2}(c_1-c_2)+\frac{\pi\lambda}{2}f c_2\right)\Big]\,,\\
&=\exp\Big[N\left(\frac{2\sqrt{\lambda}}{3\pi}\left(\sqrt{1-\nu}\sin\theta_{2}\right)^3+(1-f)\frac{k\nu\lambda}{4N}\right)\Big].
\label{antiaction1}
\end{align}

There is an additional saddle point sitting on the interval $[-\frac{1}{\mu_2}(c_1-c_2)-\frac{\mu_1}{\mu_2},-\frac{1}{\mu_2}(c_1-c_2)+\frac{\mu_1}{\mu_2}]$. In this case, the first integral on equation \eqref{antisymeom} vanishes, whereas the second one only receives contributions from $0<x<\tilde{z}$ with
\be
\tilde{z}=\frac{1}{\mu_1}\left(\mu_2z+c_1-c_2\right)\,, \quad -1<\tilde{z}<1\, ,
\ee
thus obtaining the following equation
\be
\text{arccos}(\tilde{z})-\tilde{z}\sqrt{1-\tilde{z}^2}=\pi\left(1-\frac{\lambda}{\mu_1^2}f\right)\, ,
\ee
which is solved in this other case by $\tilde{z}=\cos\theta_{1}$ such that
\begin{align}
\theta_{1}-\sin\theta_{1}\cos\theta_{1}&=\pi\left(1-\frac{\lambda}{\mu_1^2}f\right)
=\pi\left(1-\frac{l}{N \nu}\right)\,.\label{theta1}
\end{align}
The integral \eqref{antisym} evaluated at this saddle contributes as
\begin{align}
\langle W_{A_l}\rangle_{\mathbf{R}}^{(1)}&\approx\exp\big[\frac{2N}{\pi\lambda}\left(\mu_1^3\int_{-1}^{\cos\theta_1} dx\, x\sqrt{1-x^2}(-\mu_1x+c_1-c_2)+\frac{\pi\lambda}{2}f c_2\right)\Big]\,,\\
&=\exp\Big[N\left(\frac{2\sqrt{\lambda}}{3\pi}\left(\sqrt{\nu} \sin\theta_{1}\right)^3+f \frac{k(1-\nu)\lambda}{4N}\right)\Big]\,.
\end{align}
Hence, the result for the correlator from both saddle points is
\begin{align}
\langle W_{A_l}\rangle_{\mathbf{R}} \approx & \exp\left[{N\left(\frac{2\sqrt{\lambda}}{3\pi}\left(\sqrt{\nu} \sin\theta_{1}\right)^3+f \frac{k(1-\nu)\lambda}{4N}\right)}\right]\nn \\
&+\exp\left[{N\left(\frac{2\sqrt{\lambda}}{3\pi}\left(\sqrt{1-\nu}\sin\theta_{2}\right)^3+(1-f)\frac{k\nu\lambda}{4N}\right)}\right]\,.
\label{antisymfinal}
\end{align}

It is worth noting that implementing the following conjugation, $\nu\to 1-\nu$ and $l\to N-l$ in \eqref{theta2} and \eqref{theta1} we find that, $\theta_1\to \pi-\theta_2$ and $\theta_2\to \pi-\theta_1$ thus leaving \eqref{antisymfinal} invariant.

\subsubsection{Back-Reacting Wilson loops in general representations}
\label{sec:GenR}

We can go further and generalize our results \eqref{symfinal} and \eqref{antisymfinal} for correlators of Wilson loops in symmetric and anti-symmetric representations with a general large representation ${\bf R}$ dual to a genus $g$ bubbling geometry. In order to do so we have to make use of the general multi-cut eigenvalue distribution \eqref{twocutdensity} proposed previously, together with the definitions of the $\mu_i$ and $c_i$ given there.

Let us consider first the symmetric case. We deform the contour of the $z$ variable to lay over the $g+1$ branch cuts of the integrand, thus obtaining the natural generalization of integral \eqref{eq:ImWL}
\begin{align}
\langle W_{S_l}\rangle_{\mathbf{R}}\approx\text{Im}\sum_i^{g+1}\frac{\mu_i}{\pi}\int_{c_i-\mu_i}^{c_i+\mu_i}\exp\Bigg[&- \frac{2N }{\pi \lambda}\Big(\mu_i^2\int_{-1}^1\sqrt{1-x^2}\log\left(1-e^{-\mu_i(x-z)}\right) + \frac{\pi\lambda}{2}f(\mu_i z-c_i)\nn\\
& +\sum_{j\neq i}\mu_j^2\int_{-1}^1\sqrt{1-x^2}\log\left(1-e^{-\mu_jx + \mu_iz +c_j-c_i)}\right)\Big) \Bigg]\,.
\end{align}
For the $i$-th term, the saddle point is located at the left of the $i$-th branch cut, but still to the right of the $(i+1)$-th one\footnote{Provided the cuts are far away from each other, this is guaranteed.}. Thus, from the sum in the second line, only the terms with center $c_j>c_i$ contribute. In our notation, this implies $j<i$, and the saddle point equations are solved by
\bea
z_i^*=-\sqrt{1+\kappa_i^2}\,, \quad \text{with} \quad\kappa_i=\frac{\lambda f}{4 \mu_i}+\frac{1}{4\mu_i}\sum_{j<i}\mu_j^2\,.
\eea
The integral evaluated at these saddle points result
\be
\exp\Bigg[\frac{2N\mu_i^2}{\lambda } \left(\kappa_i\sqrt{1+\kappa_i ^2} + \text{arcsinh}\, \kappa_i\right)+ 4N\mu_i\kappa_ic_i-\frac{N}{\lambda}\sum_{j<i}\mu_j^2 c_j + i \phi_i  \Bigg]\,,
\ee
where $\phi_i$ denotes an irrelevant phase. Taking the imaginary part and collecting all together we obtain
\be
\langle W_{S_l}\rangle_{\mathbf{R}}\approx\sum_i^{g+1}\exp\Bigg[\frac{2N\mu_i^2}{\lambda } \left(\kappa_i\sqrt{1+\kappa_i ^2} + \text{arcsinh}\, \kappa_i\right)+ 4N\mu_i\kappa_ic_i-\frac{N}{\lambda}\sum_{j<i}\mu_j^2 c_j   \Bigg]\,.
\ee

Finally, let us now turn to the antisymmetric case. Making the change of variable $t=e^{c_{g+1}-\mu_{_{g+1}} z}$, expression \eqref{generating} can be taken to the form
\begin{align}
\langle W_{A_l}\rangle_{\mathbf{R}}\approx\int_{\tilde{\Gamma}} dz\exp\Big[& \frac{2N}{\lambda \pi}\Big( \mu_{g+1}^2 \int\limits_{-1}^{1} dx \sqrt{1-x^2}\log\left(1+ e^{-\mu_{g+1}(x-z)}\right)
\label{antigeneral}
\\
+&\sum_{i}^{g}\mu_i^2 \int\limits_{-1}^{1} dx \sqrt{1-x^2}\log\left(1+ e^{-\mu_ix+\mu_{_{g+1}}z+(c_i-c_{g+1})}\right)-\frac{\pi\lambda }{2}(\mu_{g+1}z-c_{g+1})f\Big)\Big]\,.\nn
\end{align}
As for the genus one case, the contour can be deformed to run along the real axis and the integral can be approximated by evaluating the integrand at the $g+1$ saddle points sitting at
\be
z^*_i=\Big[\frac{1}{\mu_{g+1}}(c_{g+1}-c_i-\mu_i),\frac{1}{\mu_{g+1}}(c_{g+1}-c_i+\mu_i)\Big]\,, \quad i=1,\ldots, g+1\,.
\ee
Defining $\tilde{z}^*_i=\frac{1}{\mu_i}(\mu_{g+1}z^*_i+c_i-c_{g+1})$, the solution to the saddle point equations can be written as $\tilde{z}^*_i=\cos\theta_{i}$ with $\theta_{i}$ such that
\be
\theta_{i}-\sin\theta_{i}\cos\theta_{i}=\pi\left(1+\sum_{j<i}\frac{\mu_j^2}{\mu_i^2}-\frac{\lambda}{\mu_i^2}f\right),
\label{generaltheta}
\ee
hence the result for the correlator can be written as
\be
\langle W_{A_l}\rangle_{\mathbf{R}}\approx \sum_{i}^{g+1}\exp\Big[N\left(\frac{2}{3\pi \lambda}\left(\mu_i \sin\theta_{i}\right)^3+fc_i+\sum_{j<i}\frac{\mu_j^2}{\lambda}(c_j-c_i)\right)\Big].
\label{antigeneralfinal}
\ee
Furthermore, it can be seen that the last expression is manifestly invariant under conjugation of the representation ${\bf R}$. Indeed, under conjugation $\nu_i\to\nu_{g+2-i}$ and $k_i\to k_{g+1-i}$ together with $f\to 1-f$, so from \eqref{generaltheta} it can be shown that
\be
\theta_{i}\to \pi - \theta_{g+2-i}\,,
\ee
and from the definition of the centers, it can be shown that $c_i\to -c_{g+2-i}$. This together with the property $c_i+\sum_{j>i}\nu_j(c_j-c_i)=-\sum_{j<i}\nu_j(c_j-c_i)$ shows that \eqref{antigeneralfinal} is invariant under conjugation.

\section{Conclusions}
\label{conclu}

We have found classical fundamental string solutions in the background of bubbling geometries dual to Wilson loops in large rank representations.
For a general genus $g$ background we have shown that minimal area configurations are found at the points $z=z^*$ of the Riemann surface $\Sigma$ that minimize both the area (given by the product of the dilaton  and the warping factor $e^{\Phi\over 2} f_1^2$) and the $B$-field component $b_1$. We have also found that the critical points, in the upper half-plane coordinates, are precisely located at the branch points $e_a$.

Furthermore, we have argued that $g+1$ out of the $2g+2$ solutions correspond to string configurations preserving the same symmetries and supersymmetries as the bubbling geometries. Thus, only the former have to be taken into account in the saddle point approximation that is related to the strong coupling limit of the correlator between a large representation Wilson loop and a fundamental Wilson loop.

In order to write down the explicit expressions for the corresponding on-shell actions, we have considered in great detail the case of strings in genus one backgrounds. In this case the on-shell actions display  quite a non-trivial structure, since two classical configurations contribute to the saddle point approximation.

In the case of genus one background, the dual large representation Wilson loop is characterized by a rectangular Young tableau. The matrix model computation we performed for its correlator with a fundamental Wilson loop is valid in the large-$N$ limit and requires $\frac{k\lambda}{4N}\gg 1$ as well.  Remarkably, the large $\lambda$ limit of this correlator, given in terms of a combination of two Bessel functions, was shown to be in perfect agreement with the two contributions to the string theory saddle point approximation.

In addition, the correlator of a fundamental and a generic Young tableau representation Wilson loop was similarly solved in the large-$N$ limit, provided the edges of the tableau are all size of order $N$. The resulting expression for the correlator is again given by a combination of $g+1$ Bessel functions. Finally, we went on to compute correlators of more general configurations including, for instance, a large rectangular representation with totally symmetric and totally anti-symmetric representations.

Let us close with some comments about  open problems that could be interesting complements of the results presented in this article.
Our computation for correlators between rectangular and totally symmetric/anti-symmetric representation Wilson loop provides a prediction for
probes D3 and D5 branes in the bubbling geometry background. Thus, it would be interesting to find those D-brane configurations and evaluate their on-shell actions.

 Alternatively, it would be interesting to consider the gravity picture suggested by  the product of characters formula in the field theory side, and check that each saddle point in the on-shell string action indeed coincides with a bubbling geometry of one genus higher, in a limit where one branch cut collapses.

Our work, together with the very interesting results of  \cite{Gomis:2008qa} where correlators of large Wilson loops with local operators were discussed, creates a platform for the computation of more general correlators. Following some of the development in \cite{Alday:2011pf}, it seems now feasible to tackle more complicated insertions, for example, two Wilson loops and a local operator. Clearly, one of our driving motivations has been a concrete exploration of non-conformal gauge/gravity pairs. However, we secretly hope that some thread of the beautiful integrability techniques that have been so successful in understanding the structure of three-point correlators \cite{Zarembo:2010rr,Gromov:2012vu} might still be extracted from our explicit computations.

Finally, and certainly more ambitiously, there is the question of sub-leading corrections on both sides of the correspondence. On the field theory side, there are well established techniques to go beyond the large-$N$ limit and they have been applied to the computation of Wilson loops in the context of the Gaussian matrix model \cite{Faraggi:2014tna,Liu:2017fiq,Gordon:2017dvy}; there are also techniques to explore the large $\lambda$ expansion in some cases \cite{Horikoshi:2016hds,Chen-Lin:2016kkk}. It will be instructive to extend these computations to correlators of Wilson loops. The holographic computation, although conceptually clear \cite{Faraggi:2011bb,Faraggi:2011ge,Buchbinder:2014nia}, seems more daunting at the moment.

\section*{Acknowledgments}
We are thankful to X. Chen-Lin, J. Liu, G. Silva, D. Trancanelli and S. Zhou for various discussions on closely related topics. LAPZ is partially supported by the US Department of Energy under Grant No. \ DE-SC0017808 -- {\it Topics in the AdS/CFT Correspondence: Precision tests with Wilson loops, quantum black holes and dualities}.
DHC and JAD are supported by CONICET and grants PICT 2012-0417 and PIP 0681 and PI {\it B\'usqueda de nueva F\'\i sica}. V.I. Giraldo-Rivera is grateful to the string theory group at IFLP for hospitality during the completion of this work. He is happy to thank the members of String theory group at ICTS-TIFR, for the support and encouragement. He also acknowledges  support from Simons Foundation. The work of FF  and JFM is partially supported by the MIUR PRIN Contract 2015MP2CX4-{\it Non-perturbative Aspects Of Gauge Theories And Strings}.

\appendix

\section{Probe brane limit}
\label{AppProbe}

As was mentioned in section \ref{strings}, the genus one geometry has two free parameters, $\omega_1$ and $\omega_3$, which are in turn related to the parameters of the Wilson loop representation, or alternatively to the number of D3 and D5 branes in the dual back-reacting brane configuration. In this appendix we consider the $\omega_1\to\infty$ regime, which corresponds to the collapse of one of the $[\tilde{e}_1,\tilde{e}_2]$ segments and the consequent recovering of the $AdS_5\times S^5$ geometry \cite{D'Hoker:2007fq,Benichou:2011aa}.

\begin{figure}[H]
\centering
\def\svgwidth{12cm}
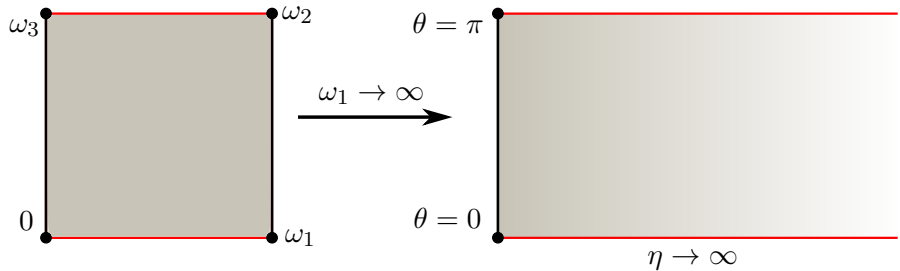
\caption{Points at $z=\{0,\omega_3\}$ are mapped to $\theta= \{0,\pi\}$ in the limit $\omega_1\to\infty$.}
\label{w1large}
\end{figure}

For this we expand the Weierstrass elliptic functions for large $\omega_1$
\begin{align}
\wp(z)&\simeq-\frac{\pi^2}{12\omega_3^2}\left(1+\frac{3}{\sinh^2\left(\frac{{\rm i}\,\pi z}{2\omega_3}\right)}\right)\,,\\
\zeta(z)&\simeq\frac{\pi^2 z}{12\omega_3^2}+\frac{{\rm i}\,\pi}{2\omega_3}\coth\left(\frac{{\rm i}\,\pi z}{2\omega_3}\right)\,.
\end{align}
In this limit, the $\omega_3$ dependence is completely artificial and does not enter in any geometrical quantity. In fact, it is possible to get rid of it by a holomorphic redefinition of the variables which, precisely for being holomorphic, does not alter the geometry. The precise form of this transformation is
\be
z=\frac{|\omega_3|}{\pi}\log\left(\frac{1+{\rm i}\, \sinh\left(\frac{\pi}{|\omega_3|}+\eta+{\rm i}\,\theta\right)}{\cosh\frac{\pi}{|\omega_3|}
+{\rm i}\,\sinh(\eta+{\rm i}\,\theta)}\right)\,,
\ee
under which the functions $h_1$ and $h_2$ become
\begin{align}
h_1&=\frac{L^2}{4\sqrt{g_S}}\cosh(\eta+{\rm i}\,\theta)+ {\rm c.c.} \,,\\
h_2&=\frac{L^2\sqrt{g_S}}{4}\sinh(\eta+{\rm i}\,\theta)+ {\rm c.c.} \,,
\end{align}
leading to the usual $AdS_2\times S^2\times S^4$ fibration metric of $AdS_5\times S^5$
\be
ds^2=L^2\left(\cosh^2{\eta}ds^2_{AdS_2}+\sinh^2{\eta}d\Omega_2^2+d\eta^2+d\theta^2+\sin^2{\theta}d\Omega_4^2\right)\,.
\label{adsmetric}
\ee
Therefore, in this limit, the fundamental domain of the Weierstrass functions is mapped to the semi-infinite strip described by $0\leq\eta<\infty$ and $0\leq\theta\leq\pi$ (see Figure \ref{w1large} ). Moreover, it is easy to see that the $z=0$ and $z=\omega_3$ are mapped to antipodal points $(\eta=0,\theta=0)$ and $(\eta=0,\theta=\pi)$, respectively.

\section{Contribution from other saddle points}

\label{sec:2ndsad}
We will now find a second saddle  that contributes for to $\langle W_{S_l}\rangle_\mathbf{R}$ in section \ref{sec:Symm}. The first integral is
 \begin{align}
 \!\!\!\frac{\mu_1}{\pi}\int\limits_{-1}^{1}\!dz & \exp\Bigg[-\frac{2N}{\pi \lambda}\Big(\mu^2_1\!\int\limits_{-1}^{z}\!dx\sqrt{1-x^2}\log(e^{\mu_1 z}-e^{\mu_1 x})+\mu^2_1\!\int\limits_{z}^{1}\!dx\sqrt{1-x^2}\log(e^{\mu_1 x}-e^{\mu_1 z})
 \\
 +{\rm i}\, & \mu_1^2\pi\!\int\limits_{-1}^{z}dx\sqrt{1-x^2}+\mu^2_2\!\int\limits_{-1}^{1}\!dx\sqrt{1-x^2}\log(1-e^{-\mu_2 x+c_2-c_1+\mu_1 z}) +\frac{\pi\lambda}{2}(\mu_1 z-c_1)f\Big) \Bigg]\,.\nonumber
 \end{align}
 We will find an additional solution to the saddle point equations proceeding as in \cite{Hartnoll:2006is}, namely taking the large $\lambda$ limit before  finding the saddle point equations. Therefore we have
 \be
-\frac{2N}{\pi \lambda}\left[\mu^3_1 z\int\limits_{-1}^{z}\!dx\sqrt{1-x^2}+\mu^3_1  \!\int\limits_{z}^{1}\!dx x \sqrt{1-x^2}+{\rm i} \,  \mu_1^2\pi\!\int\limits_{-1}^{z}dx\sqrt{1-x^2}
 +\frac{\pi\lambda}{2}(\mu_1 z-c_1)f\right]\,,
 \ee
 yielding
 \be
 \frac{\mu^3_1}{\lambda} \int\limits_{-1}^{z}\!dx\sqrt{1-x^2}+\, {\rm i}\,  \frac{\mu_1^2}{\lambda}\pi\sqrt{1-z^2} +\frac{\pi}{2}\mu_1 f\approx
 \frac{\mu^3_1}{\lambda} \int\limits_{-1}^{z}\!dx\sqrt{1-x^2}+ \frac{\pi}{2}\mu_1 f=0\,,
\ee
where in the expression in the r.h.s. we have discarded the term proportional $\frac{\mu^2_1}{\lambda}$ since it is sub-leading in the large $\lambda$ limit.
The resulting equation is completely analogous to the one found in \cite{Hartnoll:2006is}, and has complex solutions
parametrized by
\bea
\tilde{z}_1=\cos\psi_1\in\mathbb{C}\,,
\eea
with $\psi_1$ satisfying
\bea
\pi\Big(\frac{f+\nu}{\nu}\Big)=\psi_1-\cos\psi_1 \sin\psi_1\,.
\eea
The evaluation of the integral in this saddle point gives the following contribution,
\bea
\langle W_{S_l} \rangle_\mathbf{R}^{(1)}\Big|_{\rm sub}&\approx &\exp\Big(-\frac{2N}{3\pi }\sqrt{\lambda}\,\text{Re}(\sqrt{\nu} \sin \psi_1)^3+ N c_1 f\Big)\,,\nonumber\\
&=&\exp\Big(-\frac{2N}{3\pi }\sqrt{\lambda}\,\text{Re}(\sqrt{\nu} \sin \psi_1)^3+  \frac{k(1-\nu)}{4}\lambda f\Big)\,.\nonumber\\
\eea
Similarly the  second integral in Eq. \eqref{eq:ImWL}, in this approximation has a saddle point equation of the form
\bea
\frac{2}{\pi}\int_{1}^z\sqrt{1-x^2}+\frac{\mu^2_1+1}{\mu_2^2}f=\frac{2}{\pi}\int_{1}^z\sqrt{1-x^2}+\frac{f+\nu}{1-\nu}=0\,,
\eea
with solutions parameterized by the complex angle $\psi_2$ satisfying
 \bea
\pi\Big(\frac{f+1}{1-\nu}\Big)=\psi_2-\cos\psi_2\sin\psi_2\,,
\eea
therefore,
\bea
\langle W_{S_l} \rangle_\mathbf{R}^{(2)}\Big|_{\rm sub}&\approx &\exp\Big(-\frac{2N}{3\pi }\sqrt{\lambda} \,\text{Re}(\sqrt{1-\nu} \sin \psi_2)^3-N \frac{\mu^2_1}{\lambda}(c_1-c_2) + N c_2 f\Big)\,,\nonumber\\
&=&\exp\Big(-\frac{2N}{3\pi }\sqrt{\lambda}\,\text{Re}(\sqrt{1-\nu} \sin \psi_2)^3- \frac{k\nu}{4} (f+1)\lambda \Big)\,.\nonumber\\
\eea
Finally, the total contribution from these saddle points is
\begin{align}
\langle W_{S_l} \rangle_\mathbf{R}^{\text{sub}}\approx &\exp\Big(-\frac{2N}{3\pi }\sqrt{\lambda}\,\text{Re}(\sqrt{\nu} \sin \psi_1)^3+  \frac{k(1-\nu)}{4}\lambda f\Big)\nonumber\\ &\qquad\qquad\qquad\qquad\qquad\qquad+\exp\Big(-\frac{2N}{3\pi }\sqrt{\lambda}\,\text{Re}(\sqrt{1-\nu} \sin \psi_2)^3- \frac{k\nu}{4} (f+1)\lambda \Big),
\end{align}
The extension to the computation of these other contributions in the general back-reacting case is straightforward.

\section{Supersymmetric correlators}
\label{susywl}
Let us find what conditions the circular curves and the internal space orientations have to fulfill in order for the correlator in the field theory to be supersymmetric.
The supersymmetry variation of the $\mathcal{N}=4$ Wilson loop \eqref{wilsonloop} is given by \cite{Zarembo:2002an}:
\bea
\label{eq:susyWL}
\delta_{\epsilon} W_{\mathbf{R}}=\tr_{\mathbf{R}}\,P\, \int_{C}ds\bar{\Psi}(\,{\rm i}\, \Gamma^\mu \dot{x}_\mu+\rho^i n_i|\dot{x}|)\epsilon(x(s)) \, W_{\mathbf{R}}\,.
\eea
Therefore, we can say that it preserves some amount of supersymmetry if there is a solution to,
\bea
(\, {\rm i}\, \Gamma^\mu \dot{x}_\mu+\rho^i n_i|\dot{x}|)\epsilon(x(s)) =0\,,
\eea
here we use conventions of \cite{Drukker:2006ga} for Dirac matrices $\Gamma$ and $\rho$, and $\epsilon(x)$, is the most general spinor parameter generating superconformal transformations,
\bea
\epsilon(x)=\epsilon_0+ x^\mu\Gamma_\mu \epsilon_1\,,
\eea
where $\epsilon_0$ and $\epsilon_1$ are constant spinors.

For the correlator of two Wilson loops we have
\bea
&&\delta_{\epsilon}\big(W_{\mathbf{R}_1}W_{\mathbf{R}_2}\big)=\tr_{\mathbf{R}_1}\,P\, \int_{C_1}ds\bar{\Psi}(\, {\rm i}\, \Gamma^\mu \dot{x}_\mu+\rho^i n^{(1)}_i|\dot{x}|)\epsilon(x(s))W_{\mathbf{R}_2}\nonumber\\
&& \qquad\qquad\qquad\qquad\qquad\qquad +W_{\mathbf{R}_1}\tr_{\mathbf{R}_2} \,P\, \int_{C_2} ds\bar{\Psi}(\, {\rm i}\, \Gamma^\mu \dot{x}_\mu+\rho^i n^{(2)}_i|\dot{x}|)\epsilon(x(s))\,.
\eea
Therefore for this correlator to be supersymmetric we need both,
\bea
(\, {\rm i}\, \Gamma^\mu \dot{x_1}_\mu+\rho^i n^{(1)}_i|\dot{x_1}|)\epsilon(x_1(s))=0\quad\text{and}\quad(\, {\rm i}\, \Gamma^\mu \dot{x_2}_\mu+\rho^i n^{(2)}_i|\dot{x_2}|)\epsilon(x_2(s))=0\,.
\label{susyconst}
\eea
The unit vectors $n_i$ are interpreted holographically as coordinates in $S^5$ \cite{Drukker:1999zq}. We are interested in coincident $\frac{1}{2}$-BPS circular Wilson loops, but allowing the possibility for the curves to have different orientations. Thus, we consider $x_a^\mu(s)=(0, \cos s, s_a\sin s,0)$, $s_a=\pm 1$ ($a=1,2$). Furthermore we allow the possibility of operators having the same or the opposite internal space orientation, so we choose $n_{i}^{(a)}=(r_a,0,0,0,0,0)$ with $r_a= \pm 1$. For these particular choices, the supersymmetric constraints (\ref{susyconst}) become
\be
\label{eq:susycorr}
(- \, {\rm i}\, \Gamma^1 \sin s+\, {\rm i}\, s_a \Gamma^2 \cos s +r_a\rho_1 )(\epsilon_0+ \cos s\Gamma^1\epsilon_1+s_a\sin s\Gamma^2 \epsilon_1)=0\, .
\ee
It is straightfoward to see that these two equations, for $a=1,2$, are satisfied for any value of the parameter $s$ if we
one imposes
\be
-\, {\rm i}\,\Gamma_1 \epsilon_0+s_ar_a \rho_1\Gamma_2\epsilon_1=0\,,
\ee
hence, if $s_a r_a =1$, both Wilson loop operators preserve the same set of supercharges, thus leading to a supersymmetric correlator . Note that this implies, besides the obvious option, $r_1 = r_2$ and $s_1 = s_2$ for which the spatial and the internal orientations are coincident,  another possibility is given by $r_1 = -r_2$ and $s_1 = -s_2$, for which the spatial and the internal orientations are simultaneously opposite.

\bibliographystyle{JHEP}

\end{document}